\documentclass[pr, twocolumn, preprintnumbers, showpacs,footnoteadded, superscriptaddress,nofootinbib,longbibliography]{revtex4-1}

\usepackage[T1]{fontenc}
\usepackage{amssymb}
\usepackage{graphicx}
\usepackage{amsmath}
\usepackage[
      colorlinks=true,
      linkcolor=blue,
      urlcolor=blue,
      filecolor=black,
      citecolor=blue,
            ]{hyperref}
\usepackage{tensor}
\usepackage{epstopdf}
\usepackage{extarrows}
\usepackage{float}
\usepackage{amsthm}
\usepackage{color}
\usepackage{soul}

\begin{document}
\title{\boldmath On Penrose inequality in holography}	
\author{Zi-Qing Xiao}
\email{kingshaw@tju.edu.cn}
\author{Run-Qiu Yang}
\email{aqiu@tju.edu.cn}
\affiliation{Center for Joint Quantum Studies and Department of Physics, School of Science, Tianjin University, Yaguan Road 135, Jinnan District, 300350 Tianjin, P.~R.~China}
%
\begin{abstract}
The recent holographic deduction of Penrose inequality only assumes null energy condition while the weak or dominant energy condition is required in usual geometric proof. This paper makes a step toward filling up  gap between these two approaches. For planar/spherically symmetrically asymptotically Schwarzschild anti-de Sitter (AdS) black holes, we give a purely geometric proof for Penrose inequality by assuming the null energy condition. We also  point out that two naive generalizations of charged Penrose inequality are not generally true and propose two new candidates. When the spacetime is asymptotically AdS but not Schwarzschild-AdS, the total mass is defined according to holographic renormalization and depends on scheme of quantization. In this case, the holographic argument implies that the Penrose inequality should still be valid but this paper use concrete example to show that whether the Penrose inequality holds or not will depend on what kind of quantization scheme we employ.

\end{abstract}
\maketitle
\noindent
\section{Introduction}\label{intro}
In general relativity, there are many famous and universal inequalities, such as the Penrose inequality~\cite{Bray:2003ns,Mars:2009cj}, the positive mass theorem~\cite{Schon:1979rg,Schon:1981vd}, the second law of black holes~\cite{Bardeen:1973gs,Wald:1999vt} and so on. As a theoretical test, Penrose inequality is related to the establishment of cosmic censorship. Specifically, given the ADM mass/energy $M$ of a 4-dimensional asymptotically flat spacetime which contains a black hole as the initial data and denoting $A$ to be minimal area of surface enclosing the apparent horizon $\sigma$, Penrose inequality states that the spacetime's total mass $M$ should be at least $\sqrt{A/16\pi}$ and the saturation appears only when the exterior is Schwarzschild. Note that $A$ is defined as the minimal area of surface enclosing the apparent horizon $\sigma$. As pointed out by Ref.~\cite{Ben-Dov:2004lmn}, the apparent horizon area, in general, may not satisfy the Penrose inequality. Penrose's argument~\cite{Penrose:1973um} is as following: If we wait a very long time, the black hole will eventually settle down to a Kerr solution. In Kerr solution, the relationship between the black hole's mass $M_\text{kerr}$ and the area of event horizon $A_\text{ev}$ is $M_\text{kerr}\ge \sqrt{A_\text{ev}/16\pi} $. Under this evolution, the black hole's mass which is described by Bondi mass can not increase. Assuming cosmic censorship and appropriate energy conditions, the apparent horizon either lies within or coincides with the event horizon. Combining with the second law of black holes that states the area of the event horizon can't decrease, Penrose got his inequality immediately:
\begin{equation}\label{p1}
M\ge \sqrt{\frac{A}{16\pi}}.
\end{equation}
It is worth noting that the above argument is based on a lot of mathematical or physical assumptions. Although mathematicians have proven that Penrose inequality is true in some certain cases, there is no general proof for Penrose inequality(see \emph{e.g.} Refs.~\cite{10.4310/jdg/1090349428,10.4310/jdg/1090349447}).
Taking the same argument, we can also conjecture Penrose inequality for 4-dimensional asymptotically AdS spacetime~\cite{Itkin:2011ph}:
\begin{equation}\label{p2}
M\ge \left(\frac{A}{16\pi}\right)^\frac{1}{2}+\frac{1}{2\ell_{\text{AdS}}^2}\left(\frac{A}{4\pi}\right)^\frac{3}{2} ,
\end{equation}
where $M$ is the total mass/energy defined according to holographic renormalization and $A$ is the minimal area to enclose apparent horizon $\sigma$ for this asymptotically AdS spacetime. Here we have absorbed the Casimir energy~\cite{Emparan:1999pm} into the definition of total mass. When the black hole is described by Schwarzschild-AdS solution, the inequality take equal sign. Another way to rephrase this conjecture is, given the same mass, the minimal area to enclose apparent horizon is bounded by the area of AdS Schwarzschild black hole's horizon.
In holography, the bulk's geometry is dual to the two asymptotically boundary's QFT state~\cite{Witten:1998qj,Aharony:1999ti}. For each boundary's reduced density matrix, its holographic entropy is proportional to the area of black hole's apparent horizon~\cite{Ryu:2006bv,Engelhardt:2017aux}. Moreover, given same total mass $M$, Ref.~\cite{Marolf:2018ldl} shows via holography that  boundary's QFT state dual to Schwarzschild-AdS black hole has the maximum entropy.
Consequently, the AdS Penrose inequality can get from the above basic holography's argument:
\begin{equation}\label{cge}
A \le \max A=A_\text{sch}.
\end{equation}
Here $A_\text{sch}$ stands for the horizon area of Schwarzschild-AdS black hole with same total energy.
This idea was recently used by Ref.~\cite{Engelhardt:2019btp} to argue the Penrose inequality in asymptotically
AdS spacetime. We note that the Refs.~\cite{Marolf:2018ldl,Engelhardt:2019btp}  though discussed the AdS black hole of which
the cross-section of event horizon has spherical topology, their discussions are regardless the
topology of event horizon. Thus, if one follows their discussions, one shall obtain generalized
Penrose inequalities of asymptotically AdS black hole with planar or hyperbolic topologies.
For charged black holes, one might wonder that the charged generalization for Penrose inequality can be argued with Penrose's original idea. However, this does not work for the charged cases. Given the initial data mass $M$ and charge $Q$, the relation between initial total mass $M$, the final black hole mass $M_{\text{RN}}$ and the final event horizon area $A_\text{ev}$ before using the second law of black holes is
\begin{equation}\label{chargeq1}
	M\geq M_{\text{RN}}=\left(\frac{A_\text{ev}}{16 \pi}\right)^{\frac{1}{2}}+\frac{1}{2\ell_{\text{AdS}}^2}\left(\frac{A_\text{ev}}{4 \pi}\right)^{\frac{3}{2}}+\frac{Q^2}{2}\sqrt{\frac{4\pi}{A_\text{ev}}}~.
\end{equation}
here we assume that no charge can be radiated away. Assuming cosmic censorship and appropriate energy conditions , the event horizon area $A_\text{ev}$ is larger than the minimal area $A$ of surface enclosing the apparent horizon according to the second law of black holes, i.e. $A_\text{ev}\geq A$. If ``\emph{right-hand side of the inequality~\eqref{chargeq1} is a monotonically increasing function of area}'', then we would obtain the charged generalization proposed by Ref.~\cite{Itkin:2011ph}
\begin{equation}\label{chargeq1b}
	M\geq\left(\frac{A}{16 \pi}\right)^{\frac{1}{2}}+\frac{1}{2\ell_{\text{AdS}}^2}\left(\frac{A}{4 \pi}\right)^{\frac{3}{2}}+\frac{Q^2}{2}\sqrt{\frac{4\pi}{A}},
\end{equation}
Unfortunately, such monotonicity is not so obvious. Thus, one should not be surprised if the charged generalization~\eqref{chargeq1b} is broken in some cases. In fact, counterexamples of \eqref{chargeq1b} has been reported by Refs.~\cite{McCormick:2019fie,Mai:2020sac} for the case $\ell_{\text{AdS}}\rightarrow\infty$. Though the original idea of Penrose's is invalid, the logic of holographic argument proposed by Ref.~\cite{Engelhardt:2019btp} still works. If such holographic argument was really true, the minimal area $A$ would be bounded by the event horizon area of Reissner-Nordstr\"{o}m (RN) black holes if fixing the total mass $M$ and charge $Q$
\begin{equation}\label{encge}
	A(M,Q) \le A_\text{RN}(M,Q)~.
\end{equation}
This is a different generalization of Penrose inequality and was holographically argued to be true by Ref.~\cite{Engelhardt:2019btp}.
%

%
Although the recent holographic argument for Penrose inequality does not need to assume the cosmic censorship, it requires matters to satisfy the null energy condition in the bulk\footnote{Penrose inequality focus on black hole's horizon and its exterior. More precisely, the matter should satisfy the null energy condition in the black hole's exterior.}~\cite{Engelhardt:2019btp}. However, in recent years people has proved Penrose inequality in some certain cases, including the asymptotically AdS spacetimes, which requires that dominant or week energy condition~\cite{Itkin:2011ph, Lee:2015xha, Husain:2017cmj}. Both dominant and week energy conditions are stronger than the null energy condition. This forms a gap between the holographic argument and current geometric proofs on Penrose inequality in asymptotically AdS spacetime. If matters decay rapidly enough near the AdS boundary, the total mass $M$ and minimal area $A$ are geometrically well defined. The Penrose inequality in this case becomes a purely geometric inequality. Since the argument of Ref.~\cite{Engelhardt:2019btp} uses the conjecture of holography, if its conclusion is true, then it is necessary to ask: is it possible to find a purely geometric proof for AdS Penrose inequality under null energy condition without referring to the unproved conjecture of holographic principle?

According to holographic argument for charged black holes, RN black holes will have the maximum entropy so that the charged generalization is given by Eq.~\eqref{encge}. If setting the AdS radius $\ell_{\text{AdS}}$ to infinity, the charged generalization~\eqref{encge}  will following generalized Penrose inequality in asymptotically flat spacetime
\begin{equation}\label{flcg}
	\left(\frac{A}{16 \pi}\right)^{1 / 2} \leq \frac{1}{2}\left[M+\sqrt{M^{2}-Q^{2}}\right].
\end{equation}
However, there are also counterexamples~\cite{McCormick:2019fie,Mai:2020sac} for such charged generalization \eqref{flcg}. So we believe that the charged generalization~\eqref{encge} from holographic argument is not general true under null energy condition. What is the correct generalization for charged case? In addition, such counterexamples also give us enough motivation and necessity to seek purely geometric checks for the conclusions obtained from holographic principle.
So far we have assumed that matters decay rapidly enough near the AdS boundary and black holes in fact are asymptotically
Schwarzschild-AdS black holes\footnote{``Asymptotically Schwarzschild-AdS'' is stronger than "asymptotically AdS", see Ref.~\cite{shi2021regularity}}.  Both total mass~\footnote{In the following proof, we will not distinguish between ADM mass and Bondi mass since the ADM mass is equal to Bondi mass for static asymptotically Schwarzschild-AdS black holes.} $M$ and the area of minimal surface\footnote{For stationary solutions, the apparent horizon will coincide with the event horizon. So the minimal area to enclose apparent horizon $A$ is just the area of event horizon.} $A$ are determined by the bulk's geometry in this case~\cite{Ashtekar:1999jx}. If the matters do not decay rapidly enough, the spacetime may be still asymptotically AdS but not asymptotically Schwarzschild-AdS. In this case, the situation will be complicated. For instance, the existence of matter on AdS boundary will contribute to the total mass for asymptotically AdS black holes according to
holographic renormalization, see $e.g.$ Refs.~\cite{Balasubramanian:1999re,Myers:1999psa}. Since total mass obtained from
holographic renormalization is not determined by bulk geometry, the null energy condition in the bulk can still guarantee the Penrose inequality?
This paper aims to answer above questions (at least partially). For static asymptotically Schwarzschild-AdS black holes, we prove that the null energy condition can guarantee the Penrose inequality only for planar/spherical horizon geometry cases, but to guarantee the inequality for hyperbolically symmetric case we have to assume weak energy condition. A concrete counterexample is given to show the Penrose inequality is broken for hyperbolic horizon geometry under null energy condition. This implies that the conclusions of Refs.~\cite{Marolf:2018ldl,Engelhardt:2019btp} implicitly depend on the topology of event horizon, though more detailed reason is still unclear for us. For charged black hole, as we have explained that the naive generalization~\eqref{chargeq1b} and holographic version~\eqref{encge} are both incorrect. We then propose two kinds of charged generalization of Penrose inequality.
As we mentioned before, the total mass is very subtle for asymptotically AdS black holes. This paper follows the standard holographic renormalization procedure~\cite{deHaro:2000vlm,Skenderis:2002wp} and obtain the holographic mass as the total mass. Without loss of generality, we construct an asymptotically AdS black hole coupled to a scalar field to check the Penrose inequality in holography. When the source of scalar field is nonzero, the spacetime is asymptotically AdS but not asymptotically Schwarzschild-AdS. In this case, we find that the null energy condition is not enough to guarantee the Penrose inequality. Exactly speaking, whether the inequality holds or not in this case depends on what kind of quantization scheme we employ.
The organization of this paper is as follows. In section~\ref{nullprove}, given the metric ansatz for static $(d+1)$-dimensional asymptotically Schwarzschild-AdS black holes, with null energy condition, we find the null energy condition guarantee the Penrose inequality only for spherically and planar
symmetric black holes. In section~\ref{chargeprove}, we propose two types of charged generalization for Penrose inequality and prove them in static planar and spherically symmetric cases. In section~\ref{holosec}, we construct a 4-dimensional Einstein-scalar gravity and numerically check the Penrose inequality with two different quantization schemes for scalar field sector.

\section{Proof of Penrose inequality with null energy condition}\label{nullprove}
In this section, we will proposed a general version of Penrose inequality in $(d+1)$-dimensional asymptotically AdS spacetime with null energy condition and prove it under spherically/planar /hyperbolic symmetric cases. But before the general proof, we firstly revisit the Penrose inequality~\eqref{p1}\eqref{p2} in 4-dimensional spacetime and give some comments about them. In this paper, we will consider three kinds of topologies for event horizon, which are denoted by the parameter $k$. For asymptotically flat spacetime, only the spherical topology $(k=+1)$ can exist in a black hole solution. However, in asymptotically AdS spacetime, the black holes can have three topologies for the cross-section of its event horizon, $i.e.$ the spherical ($k=1$)/planar ($k=0)$/hyperbolic ($k=-1$) topologies. The Penrose inequality then should be generalized into
\begin{equation}\label{4dcase}
	M \geq\left(\frac{A}{16 \pi}\right)^{\frac{1}{2}}k+\frac{1}{2 \ell_{\mathrm{AdS}}^{2}}\left(\frac{A}{4 \pi}\right)^{\frac{3}{2}},
\end{equation}
As pointed by Ref.~\cite{Itkin:2011ph}, there is nonzero Casimir energy~\cite{Emparan:1999pm} for spherical and hyperbolic topologies. Here we have absorbed the Casimir energy into the definition of total mass to simplify our notations. For the hyperbolic and planar geometries, the volume of cross section will be infinite, which will lead the inequality~\eqref{4dcase} meaningless. However, for a static asymptotically AdS spacetime,
we can always choose coordinate gauge so that the leading term of metric near the AdS
boundary has following form
\begin{equation}
	\mathrm{d} s^{2}=-r^{2} \mathrm{d} t^{2}+\frac{\mathrm{d} r^{2}}{r^{2}}+r^{2}\mathrm{d}\Sigma_{k, d-1}^{2}~,
\end{equation}
Here$\mathrm{~d} \Sigma_{k, d-1}$ is the transverse metric of unit sphere/planar/hyperbolid defined by Eq.~\eqref{horizon}. Denote $\Omega_{k, d-1}:=\int \mathrm{d} \Sigma_{k, d-1}$. For event horizon, we can always define an ``effective'' radius $r_h$ according to equation
\begin{equation}
	A=\Omega_{k, d-1} r_{h}^{d-1}~.
\end{equation}
Similarly, we can introduce a ``mass density parameter'' $f_0$ according to
\begin{equation}\label{mass}
	M=\frac{(d-1) \Omega_{k, d-1}}{16 \pi } f_0^d
\end{equation}
Although both the total mass $M$ and area $A$ are infinite in planar or hyperbolic cases, we should notice that the mass parameter $f_0^d$ and the horizon radius $r_h$ are always finite. The Penrose inequality~\eqref{4dcase} then can be reorganized in term of following inequality (See, e.g. \cite{Itkin:2011ph, Lee:2015xha, Husain:2017cmj})
\begin{equation}\label{con1}
	\frac{1}{\ell_{\mathrm{AdS}}^{2}}+\frac{k}{r_{h}^{2}}-\frac{f_{0}^{d}}{r_{h}^{d}} \leq 0
\end{equation}
for general dimensional and all three different topologies of horizon. We then propose
following conjecture for static asymptotically Schwarzschild-AdS black holes
\newtheorem{theorem}{Conjecture}
\begin{theorem}\label{conjecture1}
	For a static asymptotically Schwarzschild-AdS black hole, if (1) Einstein equation is satisfied, (2) Matter's energy momentum tensor $T_{\mu \nu}$ satisfies null energy condition, and (3) the cross section of event horizon has spherical or planar topology, then the inequality~\eqref{con1} is true and the saturation appears only if the exterior of event horizon is Schwarzschild-AdS.
\end{theorem}
Note that the parameters $f_0$ and $r_h$ in asymptotically Schwarzschild-AdS black hole will
be determined completely by the bulk geometry, so we expect there should be a geometrical
proof without referring to the conjecture of AdS/CFT. If we recall the inequality~\eqref{con1}, the conjecture~\ref{conjecture1} then implies
\begin{equation}
	f_{0}^{d} \geq r_{h}^{d}\left(\frac{1}{\ell_{\mathrm{AdS}}^{2}}+\frac{k}{r_{h}^{2}}\right)~.
\end{equation}
Combing with the definition for total mass~\eqref{mass}, we can see that the Penrose inequality is the stronger version of the positive energy theorem, if the cross-section of event horizon
has planar or spherical topology. This is interesting and seemingly surprising since the
local energy density could be negative under null energy condition. In following, we will
give such geometrical proof in spherically/planar symmetric static cases. We will also give a detailed counterexample to show that inequality~\eqref{con1} can be broken in hyperbolic topology if we impose only null energy condition.
\subsection{Einstein equation in spherically/planar/hyperbolically symmetric cases}
For spherical/planar/hyperbolic symmetric geometries, the metric ansatz for asymptotically $(d+1)$-dimensional black holes is given by
\begin{equation}\label{as}
\mathrm{d} s^{2}=-f(r) e^{-\chi(r)} \mathrm{d} t^{2}+\frac{\mathrm{d} r^{2}}{f(r)}+r^{2}\mathrm{d}\Sigma_{k, d-1}^{2},
\end{equation}
here $k=0,\pm 1$ represents different symmetric cases
\begin{equation}\label{horizon}
d \Sigma_{k, d-1}^{2}= \begin{cases}d \Omega_{d-1}^{2}=d \theta^{2}+\sin ^{2} \theta d \Omega_{d-2}^{2}, & \text { for } k=+1 \\ d \ell_{d-1}^{2}=\sum_{i=1}^{d-1} d x_{i}^{2},  & \text { for } k=0 \\ d \Xi_{d-1}^{2}=d \theta^{2}+\sinh ^{2} \theta d \Omega_{d-2}^{2}. & \text { for } k=-1\end{cases}
\end{equation}
There is an event horizon\footnote{For a static solution, the outermost horizon will coincide
with the event horizon~\cite{Hawking:1971vc,Hawking:1973uf}.} for black hole at $r=r_h$ which is the largest root of $f(r)=0$. Outermost horizon condition will lead to $f'(r_h) \geq 0$, $i.e.$ the derivative of blackening factor $f(r)$ with respect to $r$ at the horizon $r_h$ is nonnegative. From the perspective of thermal ensemble, the temperature of black holes is nonnegative because the temperature is given by
\begin{equation}
	T=\frac{e^{-\chi(r_h)/2}}{4 \pi} f'(r_h)\geq0\,.
\end{equation}
In the following proof and following sections, we will set $\ell_{\mathrm{AdS}}=1$ for convenience. The energy momentum tensor $T\indices{^\mu_\nu}$ has a form
\begin{equation}
8 \pi T\indices{^\mu_\nu}=\operatorname{diag}\left[-\rho(r), p_{r}(r), p_{T}(r), p_{T}(r), \cdot \cdot \cdot, p_{T}(r)\right].
\end{equation}
The Einstein equation shows following three independent equations
\begin{subequations}
	\begin{align}
		f^{\prime}=& \frac{d-2}{r} k-\frac{2}{d-1}r\hat\rho-\frac{(d-2)f}{r},\label{e1} \\
		\chi^{\prime}=&-\frac{2r}{(d-1) f}\left(\rho+p_{r}\right), \label{e2}\\
		p_{r}^{\prime}=& \frac{(d-2) \hat{\rho}+2(d-1) \hat{p}_{T}-d \hat{p}_{r}}{2 r}\label{e3} \\ \notag
		&-\frac{\left(\hat{p}_{r}+\hat{\rho}\right)\left[\hat{p}_{r} r^{2}+(d-2)(d-1) k / 2\right]}{(d-1)rf} ,
	\end{align}
\end{subequations}
here
\begin{equation}
\hat{\rho}=\rho-\frac{d(d-1) }{2},~ \hat{p}_{r}=p_{r}+\frac{d(d-1)}{2},~ \hat{p}_{T}=p_{T}+\frac{d(d-1)}{2} .
\end{equation}
the extra factor $\frac{d(d-1) }{2}$ is contributed by the cosmological constant term $\Lambda g_{\mu\nu}$ in Einstein equation. As is known to all, in order to match the asymptotically AdS boundary condition, the two functions $f(r)$ and $\chi(r)$ must follow the asymptotically behaviors as
\begin{equation}\label{bou}
\lim _{r \rightarrow \infty} \frac{f(r)}{r^2}=1, \quad \lim _{r \rightarrow \infty} \chi(r)=0~.
\end{equation}
But the ``asymptotically AdS'' is not enough for our proof. Moreover, we need the matter decays rapidly near the AdS boundary $r\rightarrow\infty$, so that the functions $f(r)$ and $\chi(r)$ satisfy following $asymptotically~Schwarzschild$-AdS boundary conditions
\begin{equation}\label{fbou}
	\begin{split}
	\lim _{r \rightarrow \infty} \frac{f(r)}{r^2}=1+\frac{k}{r^{2}}-f_{0}^{d} / r^{d}+\cdots, \\
	\lim _{r \rightarrow \infty} \chi(r)=\chi_{0} / r^{d+\alpha}+\cdots, ~ \alpha>0.
	\end{split}
\end{equation}
By virtue of the asymptotic behavior for $f(r)$ and $\chi(r)$, we can define a "quasi-local mass" for an equal-$r$ surface
\begin{equation}\label{qlmass}
m(r)=\frac{k}{d}\left[r^{d-2}+X(r)\right]+\frac{r^{d+1} e^{\chi / 2}}{2 d}\left(\frac{f e^{-\chi}}{r^2}\right)^{\prime},
\end{equation}
where
\begin{equation}
X(r)=(d-2) \int_{r}^{\infty}\left[1-e^{-\chi(x) / 2}\right] x^{d-3} \mathrm{~d} x.
\end{equation}
One can check that~\footnote{For instance $k=+1, d=3$, the total mass $M$ is equal to $m(\infty)=f_{0}^{d} / 2$. This is why we call $m(r)$ "quasi-local mass".}
\begin{equation}
m(\infty)=f_{0}^{d} / 2 .
\end{equation}
We can use energy density $\rho$ and transverse pressure density $p_T$ to express $m'(r)$. One can verify
\begin{equation}\label{deri}
m^{\prime}(r)=\frac{r^{d-1} \mathrm{e}^{-\chi / 2}}{d}\left(\hat{\rho}+\hat{p}_{T}\right)=\frac{r^{d-1} \mathrm{e}^{-\chi / 2}}{d}\left(\rho+p_{T}\right) .
\end{equation}
If the matter satisfies null energy condition, combining with Eqs.~\eqref{e2} and~\eqref{deri}, we can  directly conclude that
\begin{equation}
\chi^{\prime} \leq 0, \quad m^{\prime} \geq 0
\end{equation}
and vice versa. The boundary condition~\eqref{fbou} also implies $\chi(r)\geq0$ and $X(r)\geq0$ outside the horizon, which we will use in the following proof.
Above all, $m^{\prime} \geq 0$ and $m(\infty)=f_{0}^{d} / 2$ imply that
\begin{equation}
m(r) \leq m(\infty)=f_{0}^{d} / 2 .
\end{equation}
At the horizon we have $f'(r_h) \geq 0$, so Eq.~\eqref{qlmass} implies
\begin{equation}\label{massin}
m(r_h) \geq \frac{k}{d}\left[r_{h}^{d-2}+X\left(r_{h}\right)\right].
\end{equation}
We can conclude that the "quasi-local mass" $m(r)$ is a monotonically increasing function outside the black hole, which takes the minimum value $\frac{k}{d}\left[r_{h}^{d-2}+X(r_h)\right]$ at the horizon and the maximum value $f_{0}^{d} / 2$ on the AdS boundary. Let's discuss three different horizon topologies, respectively.
\subsection{Planar Geometry}
For planar horizon case $k=0$, the expression of $m(r)$ is reduced to
\begin{equation}
m(r)=\frac{r^{d+1} e^{\chi / 2}}{2 d}\left(\frac{f e^{-\chi}}{r^2}\right)^{\prime} .
\end{equation}
Solving $f e^{-\chi}/r^2$ in terms of $m(r)$ and $\chi(r)$, we obtain
\begin{equation}\label{c1}
\frac{f e^{-\chi}}{r^2}=2 d \int_{r_{h}}^{r} \frac{m(x) {e}^{-\chi(x) / 2}}{x^{d+1}} \mathrm{~d} x,
\end{equation}
When $r\rightarrow\infty$, boundary condition indicates that
\begin{equation}
1=2 d \int_{r_{h}}^{\infty} \frac{m(x) {e}^{-\chi(x) / 2}}{x^{d+1}} \mathrm{~d} x,
\end{equation}
If the null energy condition is satisfied,  then we have $0\leq{e}^{-\chi(r) / 2}\leq 1$. Combined with $0\leq m(r)\leq f_{0}^{d} / 2$ in planar case, the above equation becomes an inequality
\begin{equation}
1=2 d \int_{r_{h}}^{\infty} \frac{m(x) {e}^{-\chi(x) / 2}}{x^{d+1}} \mathrm{~d} x \leq d \int_{r_{h}}^{\infty} \frac{f_{0}^{d}}{x^{d+1}} \mathrm{~d}x=f_{0}^{d} / r_{h}^{d},
\end{equation}
which is the Penrose inequality~\eqref{con1} for the planar symmetric case.
The inequality is saturated only if all unequal signs take the equal sign, which means $\chi=0$ and $m=f_{0}^{d} / 2$. From Eq.~\eqref{c1}, we can solve $f(r)$ in terms of $\chi(r)$ and $m(r)$ which  both are constant in this case. The solution is
\begin{equation}
	f(r)=r^2(1-f_{0}^{d}/r^d),~~\rho=p_r=p_T=0\,,
\end{equation}
which is exactly the metric of Schwarzschild-AdS black hole. Thus, we conclude that: for planar symmetric static asymptotically Schwarzschild-AdS back holes, if the null energy condition is satisfied, then Penrose inequality is true
and its saturation appears only if the black hole is Schwarzschild-AdS black hole.
\subsection{Spherical Geometry}
In this subsection, we will consider spherical symmetric $(k=1)$ case. We still firstly solve function $f(r)$ in terms of $m(r)$ and $\chi(r)$ and obtain
\begin{equation} \label{fr}
\frac{f(r){e}^{-\chi(r)}}{r^2} =2\int_{r_{h}}^{r} \frac{\left[d m(y)-X(y)-y^{d-2}\right]{e}^{-\chi(y) / 2}}{y^{d+1}} \mathrm{~d} y ,
\end{equation}
When $r$ evolves to $\infty$, the left hand of Eq.~\eqref{fr} becomes unit one
\begin{equation}
1=2 \int_{r_{h}}^{\infty} \frac{\left[d m(x)-X(x)-x^{d-2}\right] {e}^{-\chi(x) / 2}}{x^{d+1}} \mathrm{~d} x .
\end{equation}
Similar to planar symmetric case, we will focus on the integral on the right hand of Eq.~\eqref{fr}.
From Eq.~\eqref{massin}, we find $m(r_h)\geq0$ because $X(r)\geq0$. Combining it with  $ m(r) \leq m(\infty)=f_{0}^{d} / 2$, then we obtain
\begin{equation} \label{int1}
1 \leq 2 \int_{r_{h}}^{\infty} \frac{\left(d f_{0}^{d} / 2-x^{d-2}\right) {e}^{-\chi(x) / 2}}{x^{d+1}} \mathrm{~d} x .
\end{equation}
Let $r_0$ to be the root of $d f_{0}^{d} / 2-x^{d-2}=0$, the condition $\chi'\leq0$ insures
\begin{equation}
\left(d f_{0}^{d} / 2-r^{d-2}\right)\left[{e}^{-\chi(r) / 2}-{e}^{-\chi\left(r_{0}\right) / 2}\right] \leq 0
\end{equation}
which leads to
\begin{equation}
\begin{split}
&\int_{r_{h}}^{\infty} \frac{\left(d f_{0}^{d} / 2-x^{d-2}\right) {e}^{-\chi(r) / 2}}{x^{d+1}} \mathrm{~d} x\\
\leq &{e}^{-\chi\left(r_{0}\right) / 2} \int_{r_{h}}^{\infty} \frac{d f_{0}^{d} / 2-x^{d-2}}{x^{d+1}} \mathrm{~d} x\,.
\end{split}
\end{equation}
Combing this result in Eq.~\eqref{int1}, then yields
\begin{equation}
1 \leq {e}^{\chi\left(r_{0}\right) / 2} \leq \int_{r_{h}}^{\infty} \frac{d f_{0}^{d}-2 x^{d-2}}{x^{d+1}} \mathrm{~d} x=\frac{f_{0}^{d}}{r_{h}^{d}}-\frac{1}{r_{h}^{2}}\,.
\end{equation}
and Penrose inequality~\eqref{con1} follows. The inequality is saturated only if $\chi=0$ and $m=f_{0}^{d} / 2$, which leads to
\begin{equation}
	f(r)=r^2(1+1/r^2-f_{0}^{d}/r^d),~\rho=p_r=p_T=0
\end{equation}
Thus, we conclude that: static asymptotically Schwarzschild-AdS back hole with spherical symmetry, if null energy condition is satisfied, then the Penrose inequality is true and its saturation appears only if the black hole is Schwarzschild-AdS black hole. We then have proved the Conjecture 1 in the spherically and planar symmetrical cases.
\subsection{Broken Case: Hyperbolic Geometry }
\begin{figure}
	\centering
	\includegraphics[width=0.4\textwidth]{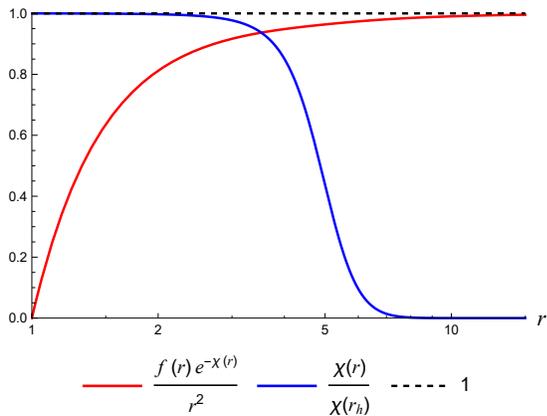}
	\caption{The values of $f(r)e^{-\chi(r)}/r^2$ and $\chi(r)/\chi(r_h)$. We can see that $f(r)e^{-\chi(r)}/r^2$ and $\chi(r)$ have the expected asymptotic behavior, but the inequality is still broken.}
	\label{fv}
\end{figure}

For hyperbolic symmetric case, the null energy condition can not guarantee Penrose inequality~\eqref{con1}. To verify this conclusion, we will give a concrete counterexample\footnote{In view of \cite{Chrusciel:2021ufc}, the validity of the Penrose inequality in general hyperbolic cases seems to be rather unlikely.}. We note that Eq.~\eqref{fr} becomes
\begin{equation}
\frac{f(r){e}^{-\chi(r)}}{r^2} =2\int_{r_{h}}^{r} \frac{\left[d m(y)+X(y)+y^{d-2}\right]{e}^{-\chi(y) / 2}}{y^{d+1}} \mathrm{~d} y .
\end{equation}
Now let us take $r_h=1, d=3$ and
\begin{equation}\label{ct}
{e}^{-\chi / 2}=\frac{4+\tanh (r-5)}{5}, \quad m(r)=f_{0}^{d} / 2 \approx-0.15105~\footnote{For hyperbolic black holes, the mass parameter $f_{0}^{d}$ can take the negative value.},
\end{equation}
we can get the expression of $X(r)$
\begin{equation}
X(r)=\frac{1}{5}[\ln \cosh (r-5)-r+\ln 2]+1 ,
\end{equation}
and two functions $f(r)e^{-\chi(r)}/r^2$ and $\chi(r)$ which are shown in Fig.~\ref{fv}. Since $\chi'< 0$ and $m'=0$, the null energy condition is guaranteed. However, let us check the sign of $\left[1-\frac{1}{r_{h}^{2}}-\frac{f_{0}^{d}}{r_{h}^{3}}\right]$, we find that
\begin{equation}
1-\frac{1}{r_{h}^{2}}-\frac{f_{0}^{d}}{r_{h}^{3}}=-\frac{f_{0}^{d}}{r_{h}^{3}}>0 .
\end{equation}
So we can conclude that the Penrose inequality~\eqref{con1} is broken when $k=-1$.
Null energy condition does not require the energy density $\rho$ nonnegative but the sum of energy density and pressure density nonnegative. If matter satisfies the weak energy condition, it can be proved that the inequality is true and its saturation appears only in the Schwarzschild-AdS black hole at least for the maximally symmetric and static cases. See appendix \ref{appendix A} for details.

\section{Charged Black Holes}\label{chargeprove}
Before we discuss the charged generalization of AdS Penrose inequality, we first consider the asymptotically charged flat case, which can be regarded as the limit of $\ell_{\text{AdS}}\rightarrow\infty$. With the total mass $M$ and charge $Q$ as initial data, there are two naive charged generalizations in 4-dimensional spacetime
\begin{equation}\label{chargeds1}
	M\geq \sqrt{\frac{A}{16 \pi}}+Q^2\sqrt{\frac{\pi}{A}}
\end{equation}
and a weaker version
\begin{equation}\label{charged}
	\left(\frac{A}{16 \pi}\right)^{1 / 2} \leq \frac{1}{2}\left[M+\sqrt{M^{2}-Q^{2}}\right].
\end{equation}
here the saturation appears only if the black holes are RN black holes. Hold on, when we introduce the charge $Q$ as a initial data as well as $M$, the definition of $Q$ is vague. For general charged black holes, $Q(r)$ is defined as
\begin{equation}\label{Q}
	\frac{1}{2\Omega_{k, d-1}} \int_{S_{r}} F_{\mu \nu} \mathrm{d}S^{\mu \nu}=Q(r),
\end{equation}
$S_{r}$ is an equal-$r$ surface and $\Omega_{k, d-1}$ denotes the dimensionless volume of the relevant horizon geometry~\eqref{horizon}.
For RN black holes, charge $Q(r)$ is a constant outside the black holes. This is because there is no charge outside RN black holes. For general cases, $Q(r)$ is dependent on $r$, because matters outside the black hole usually also carry charge. If interpreting $Q^2$ in inequality~\eqref{charged} as the square of total charge, $i.e. ~Q^2(\infty)$, we shall find that the inequality~\eqref{charged} is not always true. See Refs.~\cite{McCormick:2019fie,Mai:2020sac} for a counterexample. This reminds us that the charged generalization for Penrose inequality needs to be treated carefully and naive generalizations~\eqref{chargeq1b} and \eqref{encge} are both incorrect in general. In this section, we will conjecture two different types of charged generalization for Penrose inequality.
\subsection{The First Type of Generalization}\label{first}
\begin{figure}[H]
	\centering
	\includegraphics[width=0.45\textwidth]{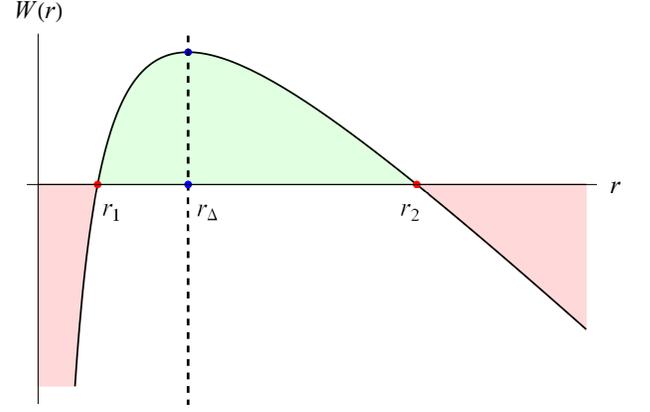}
	\caption{The schematic diagram of function $W(r)$ for $k=1$. The value of function $W(r)$ at the horizon $r_h$ must be nonnegative.}
	\label{wr}
\end{figure}

We will separate the energy momentum tensor $T_{\mu\nu}$ into two parts,
\begin{equation}
	T_{\mu \nu}=T_{\mu \nu}^{(M)}+T_{\mu \nu}^{(o)}
\end{equation}
where the energy momentum tensor $T_{\mu \nu}^{(M)}$ for Maxwell field is defined as
\begin{equation}
	T_{\mu \nu}^{(M)}=2\left(F\indices{_\mu^\sigma}F_{\nu \sigma}-\frac{g_{\mu \nu}}{4} F_{\sigma \tau} F^{\sigma \tau}\right) ,
\end{equation}
and $T_{\mu \nu}^{(o)}$ stands for other parts in $T_{\mu \nu}$. To present our generalized Penrose inequality in static charged black holes, we needs a little more preparation.
Denote $\gamma$ to be a co-dimensional-2 spacelike surface. Denote that $l^\mu$ to be future-directed infalling null geodesic vectors
field which are normal to $\gamma$ and satisfies $\xi^\mu l_\mu=-1$, where $\xi^\mu$ is the Killing vector standing for static symmetry and normalized at infinity by $\xi^\mu \xi_\mu=-1$. We denote the expansion $\theta_{(l)}$ for $l^\mu$. Then we define $Q_m^2$ to be
\begin{equation}\label{QM}
	Q_{m}^2=\inf _{S}\left[\frac{(1-d)Q^2(\gamma)S(\gamma)}{r_s\int_\gamma\theta_{(l)}\mathrm{d}S}\right].
\end{equation}
$r_s$ is an "effective" radius which satisfies
\begin{equation}
	r_s^{d-1}\Omega_{k, d-1}=S(\gamma)~,
\end{equation}
here $S(\gamma)$ denotes the area of $\gamma$. We now propose the first type of charged generalization:
\newtheorem{theorem1}{Theorem}
\begin{theorem}
	For an asymptotically Schwarzschild-AdS black hole, if (1) Einstein equation is satisfied, (2) $T_{\mu \nu}^{(o)}$ satisfies null energy condition, and (3) the cross section of event horizon has spherical or planar topology, then the charged generalization of Penrose inequality reads
	\begin{equation}\label{cpen}
		1+\frac{k}{r_{h}^{2}}-\frac{f_{0}^{d}}{r_{h}^{d}}+\frac{2Q_m^2}{(d-1)(d-2)r_h^{2d-2}} \leq 0~.
	\end{equation}
The saturation appears only in RN black holes.
\end{theorem}
This one is very similar to the generalization proposed by Ref.~\cite{Itkin:2011ph}, however,  $Q_m$ here in general will be different from the total charge.
To support our this generalization, we will given the proof for spherically and planar symmetric cases. Under coordinates gauge~\eqref{as}, the expansion $\theta_{(l)}$ for $l^\mu$ is given by
\begin{equation}
	\theta_{(l)}=(1-d)\frac{e^{\chi/2}}{r}
\end{equation}
and the Maxwell field strength tensor has a form
\begin{equation}\label{maxfield}
	F_{\mu \nu}=-\frac{Q(r) e^{-\chi / 2}}{r^{d-1}}(\mathrm{d} t)_{\mu} \wedge(\mathrm{d} r)_{\nu}.
\end{equation}
The nonvanishing components of surface element $dS_{\mu\nu}$ read
\begin{equation}
	dS_{01}=-dS_{10}=e^{-\chi/2}r^{d-1}\mathrm{d}\Sigma_{k, d-1}.
\end{equation}
Substituting this result into the definition~\eqref{QM}, we can obtain the expression of $Q^2_m$
\begin{equation}\label{Qm}
	Q^2_m=\min\left[Q^2(r)e^{-\frac{\chi(r)}{2}}\right],~\text{for}~r\geq r_h
\end{equation}
In order to prove this inequality~\eqref{cpen}, the key step is to separate the energy density $\rho$ and pressure density $\left\{p_r,p_T\right\}$ into two parts respectively
\begin{equation}\label{sep}
	\rho=\frac{Q^{2}}{ r^{2d-2}}+\rho^{(o)}, \quad p_{r}=-\frac{Q^{2}}{ r^{2d-2}}+p_{r}^{(o)}, \quad p_{T}=\frac{Q^{2}}{r^{2d-2}}+p_{T}^{(o)}.
\end{equation}
Since we require that $\rho^{(o)}+p_{T}^{(o)} \geq 0$ and $\rho^{(o)}+p_{r}^{(o)} \geq 0$, then we obtain that
\begin{equation}
	\rho+p_{r} \geq 0~, \quad \rho+p_{T} \geq \frac{2Q^{2}}{ r^{2d-2}}~.
\end{equation}
To prove the inequality~\eqref{cpen}, we introduce a new "quasi-local mass" $\tilde{m}(r)$
\begin{equation}\label{new}
\tilde{m}(r)=m(r)+\int_{r}^\infty\frac{2Q^{2}e^{-\frac{\chi}{2}}}{d y^{d-1}} \mathrm{~d} y~,
\end{equation}
 so that the derivative of $\tilde{m}(r)$ is always nonnegative
\begin{equation}\label{}
	\tilde{m}^{\prime}(r)=\frac{r^{d-1} \mathrm{e}^{-\chi / 2}}{d}\left(\rho+p_{T}-\frac{2Q^{2}}{ r^{2d-2}}\right)\geq 0~.
\end{equation}
and when $r\rightarrow\infty$
\begin{equation}\label{}
	\tilde{m}(\infty)=m(\infty)=d f_{0}^{d} / 2~.
\end{equation}
This implies
\begin{equation}\label{maxqm}
	\tilde{m}(r)\leq d f_{0}^{d} / 2~.
\end{equation}
Recall the definition of old "quasi-local mass"~\eqref{qlmass}, we substitute the expression of $m(r)$ into our new "quasi-local mass"~\eqref{new}
\begin{equation}
\begin{split}
	\tilde{m}(r)&=\frac{k}{d}\left[r^{d-2}+X(r)\right]+\frac{r^{d+1} e^{\chi / 2}}{2 d}\left(\frac{f e^{-\chi}}{r^2}\right)^{\prime}\\
&+\int_{r}^\infty\frac{2Q^{2}e^{-\frac{\chi}{2}}}{d y^{d-1}} \mathrm{~d} y,
\end{split}
\end{equation}
Here $k=0$ and $1$, which stands for planar and spherically symmetry respectively. Following the standard procedure in section~\ref{nullprove}, we should solve $f e^{-\chi}/r^2$ in terms of $m(r)$ and $\chi(r)$
\begin{equation}\label{ordi}
	\begin{split}
	&2\int_{r_{h}}^{r} \left[d \tilde{m}(x)-\int_{x}^\infty\frac{2Q^{2}e^{-\frac{\chi}{2}}}{y^{d-1}} \mathrm{~d} y-kX(x)\right.\\
&-kx^{d-2}]{e}^{-\chi(x) / 2}x^{-(d+1)} \mathrm{~d} x=\frac{f(r){e}^{-\chi(r)}}{r^2},
	\end{split}
\end{equation}
Recall that $f(r_h)'\geq0$ because the surface of $r=r_h$ is outermost horizon, so we can obtain
\begin{equation}\label{rh}
\frac{\left[d \tilde{m}(r_h)-\int_{r_h}^\infty\frac{2Q^{2}e^{-\frac{\chi}{2}}}{y^{d-1}} \mathrm{~d} y-kX(r_h)-kr_h^{d-2}\right]{e}^{-\chi(r_h) / 2}}{r_h^{d+1}}\geq0~.
\end{equation}
Due to $X(r)\geq0$ and ${e}^{-\chi(r_h) / 2}\geq0$, the above inequality~\eqref{rh} becomes
\begin{equation}\label{inrh}
	\frac{d \tilde{m}(r_h)-\int_{r_h}^\infty\frac{2Q^{2}e^{-\frac{\chi}{2}}}{y^{d-1}} \mathrm{~d} y-kr_h^{d-2}}{r_h^{d+1}}\geq0~.
\end{equation}
We shall see that, the above inequality which is defined at the horizon $r_h$ plays a decisive role in the following proof. In particular, the inequality~\eqref{inrh} restrict the evaluation relationship between total mass $M$ and $Q_m^2$. For the convenience of our proof, we define an auxiliary function $W(r)$
\begin{equation}
 W(r)=d f_{0}^{d} / 2-\frac{2Q_m^2}{(d-2)r^{d-2}}-kr^{d-2}.
\end{equation}
Combine it with the Eqs.~\eqref{Qm},~\eqref{maxqm} and~\eqref{inrh} and we will obtain
\begin{equation}\label{W(r)}
	\frac{W(r)}{r^{d+1}}\geq\frac{d \tilde{m}(r)-\int_{r}^\infty\frac{2Q^{2}e^{-\frac{\chi}{2}}}{y^{d-1}} \mathrm{~d} y-kr^{d-2}}{r^{d+1}}~.
\end{equation}
Particularly, at horizon $r=r_h$ we have
\begin{equation}
	\frac{W(r_h)}{r_h^{d+1}}\geq\frac{d \tilde{m}(r_h)-\int_{r_h}^\infty\frac{2Q^{2}e^{-\frac{\chi}{2}}}{y^{d-1}} \mathrm{~d} y-kr_h^{d-2}}{r_h^{d+1}}\geq 0~,
\end{equation}
due to inequality~\eqref{inrh}. Thus the horizon $r_h$ is limited by the value of function $W(r)$.
Back to Eq.~\eqref{ordi}, the left hand of Eq.~\eqref{ordi} will become unit one when $r$ evolves to $\infty$, which is the boundary condition of asymptotically AdS spacetime
\begin{equation}
\begin{split}
	1=&2\int_{r_{h}}^{\infty}\left[d \tilde{m}(x)-\int_{x}^\infty\frac{2Q^{2}e^{-\frac{\chi}{2}}}{y^{d-1}} \mathrm{~d} y-kX(x)\right.\\
&-kx^{d-2}]{e}^{-\chi(x) / 2}x^{-(d+1)} \mathrm{~d} x~.
\end{split}
\end{equation}
Through Eqs.~\eqref{massin}~\eqref{new}, we find $\tilde{m}(r_h)\geq0$. Combining it with   $\tilde{m}(r) \leq \tilde{m}(\infty)=f_{0}^{d} / 2$ and $X(r)\geq0$, we then obtain
\begin{equation}
	1 \leq 2\int_{r_{h}}^{\infty} \frac{\left[d f_{0}^{d} / 2-\int_{x}^\infty\frac{2Q^{2}e^{-\frac{\chi}{2}}}{y^{d-1}} \mathrm{~d} y-kx^{d-2}\right]{e}^{-\chi(x) / 2}}{x^{d+1}} \mathrm{~d} x~.
\end{equation}
Using the inequality~\eqref{W(r)}, we see that the above inequality becomes
\begin{equation}\label{intcharge}
	1 \leq 2\int_{r_{h}}^{\infty} \frac{W(x){e}^{-\chi(x) / 2}}{x^{d+1}} \mathrm{~d} x.
\end{equation}
Above inequality implies that the maximum value of $W(r)$ must be positive
\begin{equation}
\max W(r)=\max \left[d f_{0}^{d} / 2-\frac{2Q_m^2}{(d-2)r^{d-2}}-kr^{d-2}\right] > 0~,
\end{equation}
otherwise the above integration Eq.~\eqref{intcharge} will be negative.
For $k=1$, it's obvious that when $r^{d-2}=r_\Delta^{d-2}=\sqrt{\frac{2Q^2_m}{d-2}}$, $W(r)$ takes the maximum value $d f_{0}^{d}/2-2\sqrt{\frac{2Q^2_m}{d-2}}$
\begin{equation}
	\max W(r)=W(r_\Delta)=d f_{0}^{d}/2-2\sqrt{\frac{2Q^2_m}{d-2}} > 0~.
\end{equation}
There are two points $r_1,r_2$ which are the roots of $W(r)=0$,
\begin{equation}
	\begin{split}
	r_1^{d-2}=d f_{0}^{d}/4-\sqrt{(d f_{0}^{d}/4)^2-\frac{2Q^2_m}{d-2}},~\\r_2^{d-2}=d f_{0}^{d}/4+\sqrt{(d f_{0}^{d}/4)^2-\frac{2Q^2_m}{d-2}}~.
	\end{split}
\end{equation}
 As we can see that $r_1\leq r_h< r_2$  from Fig.~\ref{wr} because $W(r_h)\geq0$~\footnote{The value of $r_h$ can not equal or greater than $r_2$, otherwise the integration of Eqs.~\eqref{intcharge} will be negative.}.
We separate the interval $[r_h,\infty)$ into two parts $[r_h,r_2)$ and $[r_2,r_\infty)$. Then there are two different situations.
\begin{itemize}
	\item $r_h\leq r<r_2$
	
	Then  $\chi'\leq0$ insures ${e}^{-\chi(r) / 2}-{e}^{-\chi(r_2) / 2}\leq0$ and $W(r)\geq0$. We then have
	\begin{equation}\label{ieq}
		W(r)\left[{e}^{-\chi(r) / 2}-{e}^{-\chi\left(r_2\right) / 2}\right] \leq 0 ,
	\end{equation}
\item $r_2\leq r$
	
	We see ${e}^{-\chi(r) / 2}-{e}^{-\chi(r_2) / 2}\geq0$ and $W(r)\leq0$, so still have inequality~\eqref{ieq}.
\end{itemize}
We see that inequality~\eqref{ieq} is true for all $r\in[r_h,\infty)$. This leads to
	\begin{equation}
		1\leq2\int_{r_{h}}^{\infty} \frac{W(r) {e}^{-\chi(r) / 2}}{r^{d+1}} \mathrm{~d} r
		\leq 2{e}^{-\chi\left(r_2\right) / 2} \int_{r_{h}}^{\infty} \frac{W(r)}{r^{d+1}} \mathrm{~d} r\,.
	\end{equation}
 Multiplying ${e}^{-\chi\left(r_2\right) / 2}$ into above inequality, we finally obtain
	\begin{equation}
\begin{split}
		1 \leq {e}^{\chi\left(r_2\right) / 2} &\leq 2\int_{r_{h}}^{\infty} \frac{W(r)}{x^{d+1}} \mathrm{~d} x\\
&=\frac{f_{0}^{d}}{r_{h}^{d}}-\frac{1}{r_{h}^{2}}-\frac{2Q_m^2}{(d-1)(d-2)r_h^{2d-2}}~.
\end{split}
	\end{equation}

For $k=0$, $W(r)$ is a monotonically increasing function and so that
\begin{equation}
	\max W(r)=W(r\rightarrow \infty)=d f_{0}^{d}/2> 0~.
\end{equation}
There is a point $r_0$ satisfies $W(r_0)=0$
\begin{equation}
	r_0^{d-2}=\frac{4Q_m^2}{d(d-2)f_0^d}~.
\end{equation}
Combining $W(r_h)\geq0$ with $W'(r)\geq0$ for planar case, we obtain that $W(r)\geq0$ for $r\geq r_h$. The inequality~\eqref{intcharge} becomes
\begin{equation}
	\begin{split}
	1 \leq 2\int_{r_{h}}^{\infty} \frac{W(x){e}^{-\chi(x) / 2}}{x^{d+1}} \mathrm{~d} x\leq2\int_{r_{h}}^{\infty} \frac{W(x)}{x^{d+1}} \mathrm{~d} x\\=  \frac{f_{0}^{d}}{r_{h}^{d}}-\frac{2Q_m^2}{(d-1)(d-2)r_h^{2d-2}}.
	\end{split}
\end{equation}
Combined two symmetric cases, the first type of charged generalization~\eqref{cpen} for Penrose inequality is followed.

Recall the whole proof, the saturation for charged inequality appears if $\chi(r)=0, \tilde{m}(r)=f_{0}^{d}/2$ and $Q(r)=Q_m$. This implies charged density $j(r)=0$ and the exterior is a RN black hole.
\subsection{The Second Type of Generalization}
The inequality~\eqref{cpen} is not expressed in term of the boundary quantities of asymptotically AdS
spacetime. It would be more satisfactory if we could use boundary quantities to express the
charged generalization of Penrose inequality since such version of Penrose inequality can
be interrupted as the inequality of dual boundary field theory according to holography. In
order to alleviate the contradiction between charged generalization for Penrose inequality
and the basic idea from the holography, this paper will propose the second type of charged Penrose inequality. Considering the Maxwell equation with source $J_\nu$
\begin{equation}
	\nabla^{\mu} F_{\mu \nu}=4 \pi J_{\nu}~.
\end{equation}
We introduce the gauge potential which satisfies
\begin{equation}
	F_{\mu \nu}=(\mathrm{d}A)_{\mu\nu}~.
\end{equation}
To find the generation in general case, we will separate the energy momentum tensor $T_{\mu \nu}^{(o)}$ into following form
\begin{equation}
	T_{\mu \nu}^{(o)}=\tilde{T}_{\mu \nu}^{(o)}-8\pi \left(g_{\mu \nu} J_\rho A^\rho-\frac{1}{2}J_{\mu} A_\nu-\frac{1}{2}J_{\nu} A_\mu\right)~,
\end{equation}
so the total energy momentum tensor reads
\begin{equation}
	T_{\mu \nu}=T_{\mu \nu}^{(M)}+\tilde{T}_{\mu \nu}^{(o)}-8\pi \left(g_{\mu \nu} J_\rho A^\rho-\frac{1}{2}J_{\mu} A_\nu-\frac{1}{2}J_{\nu} A_\mu\right)~,
\end{equation}
In static case, the gauge potential $A_\mu$ and charge density $J_\mu$ have following form
\begin{equation}
	A_{\mu} \propto \xi_{\mu}, \quad J_{\mu} \propto \xi_{\mu}~.
\end{equation}
here the potential $\Phi:=A_\mu\xi^\mu$ and charged density $j:=J_\mu\xi^\mu$. In holography, the potential $\Phi_\infty$ which is defined on the AdS boundary\footnote{In this paper, we abbreviate $\Phi(\infty)$ as $\Phi_\infty$.} is interpreted as chemical potential. Given the initial data $f_{0}^{d}/2$ and $\Phi_\infty$ and take the gauge $\Phi=0$ at event horizon, we have the following conjecture
\begin{theorem}
	For an asymptotically Schwarzschild-AdS black hole, if (1) Einstein equation is satisfied, (2) $\tilde{T}_{\mu \nu}^{(o)}$ and ${T}_{\mu \nu}$ are both satisfy null energy condition, (3) charge of black hole and charge density $j$ have same sign, and (4) the cross section of event horizon has spherical or planar topology, then the charged generation of Penrose inequality reads
	\begin{equation}\label{second}
		1 \leq \frac{f_{0}^{d}}{r_{h}^{d}}-\frac{k}{r_{h}^{2}}-\frac{2(d-2)\Phi_\infty^2}{(d-1)r_h^2}
	\end{equation}
and the saturation appears only if the exterior of event horizon is AdS-RN.
\end{theorem}
Differing from naive generalizations~\eqref{chargeq1b} and \eqref{encge}, here we use chemical potential to replace the charge. Like previous sections, to support this conjecture we will prove it under the
spherically or planar symmetric spacetime.

Under the same coordinates gauge~\eqref{as}, the gauge potential $A$ and charge density $J_\mu$ have following form
\begin{equation}
	A_{\mu}=\Phi(r)(\mathrm{d} t)_{\mu}, \quad J_{\mu}=j(r)(\mathrm{d} t)_{\mu}~.
\end{equation}
According to Eq.~\eqref{maxfield}, we obtain
\begin{equation}\label{rel}
	\frac{Q e^{-\chi / 2}}{r^{d-1}}=\Phi^{\prime} .
\end{equation}
The Maxwell equation reads
\begin{equation}\label{max}
	\left(\Phi^{\prime} e^{\chi / 2} r^{d-1}\right)^{\prime}=\frac{4 \pi j e^{\chi / 2} r^{d-1}}{f}.
\end{equation}
The energy density $\rho$ and pressure density $\left\{p_r,p_T\right\}$ now is replaced by
\begin{equation}
	\begin{split}
	\rho=\Phi^{\prime 2} e^{\chi}+\tilde{\rho}^{(o)}\qquad\qquad~~~~~, \\\quad p_{r}=-\Phi^{\prime 2} e^{\chi}+\frac{8\pi\Phi j e^{\chi}}{f}+\tilde{p}_{r}^{(o)}, \\\quad p_{T}=\Phi^{\prime 2} e^{\chi}+\frac{8\pi\Phi j e^{\chi}}{f}+\tilde{p}_{T}^{(o)}~~.
	\end{split}
\end{equation}
Null energy condition requires $\rho+ p_{r}\geq0,~\rho+ p_{T}\geq0$, then we obtain
\begin{equation}\label{nullcondition}
	\frac{8\pi\Phi j e^{\chi}}{f}+\tilde{\rho}^{(o)}+\tilde{p}_{r}^{(o)}\geq0,~2\Phi^{\prime 2}+\frac{8\pi\Phi j e^{\chi}}{f} +\tilde{\rho}^{(o)}+\tilde{p}_{T}^{(o)}\geq0
\end{equation}
A new "quasi-local mass" $\tilde{m}(r)$ is defined as
\begin{equation}
	\tilde{m}(r)=m(r)-\frac{2}{d}\left(\Phi Q\right)~,
\end{equation}
so that the derivative of $\tilde{m}(r)$ is
\begin{equation}
	\begin{aligned}
		\tilde{m}^{\prime}(r)&=m'(r)-\frac{2}{d}\left(\Phi\Phi' e^{\chi/2}r^{d-1}\right)'\\
		&=m'(r)-\frac{2}{d}e^{\chi/2}r^{d-1}\Phi'^2-\frac{2}{d}\Phi\left(e^{\chi/2}r^{d-1}\Phi'\right)'~.
	\end{aligned}
\end{equation}
Substituting Eq.\eqref{deri}, Eq.\eqref{nullcondition} and Eq.\eqref{max} into above equation and we obtain
\begin{equation}\label{ordinarymatter}
	\tilde{m}^{\prime}(r)=\frac{r^{d-1} \mathrm{e}^{-\chi / 2}}{d}\left(\tilde{\rho}^{(o)}+\tilde{p}_{T}^{(o)}\right)\geq0
\end{equation}
So we obtain $\tilde{m}(r)\leq\tilde{m}(\infty)=f_{0}^{d} / 2-\frac{2}{d}(\Phi_\infty Q_\infty)$. Let us rephrase the Maxwell equation
\begin{equation}
	Q'=\frac{4 \pi j e^{\chi / 2} r^{d-1}}{f}~.
\end{equation}
Because $Q(r_h)$ and $j$ have same sign, we can take $Q(r_h)\geq0$ and $j\geq0$ without losing generality and so charge $Q(r)$ will always nonnegative
\begin{equation}
	Q(r)\geq0~.
\end{equation}
According to the relationship between $Q$ and $\Phi$~\eqref{rel}, we can obtain
\begin{equation}
	\Phi'\geq0~.
\end{equation}
In holography, we generally set the value of potential $\Phi$ at the horizon equal to zero as a gauge fixing
\begin{equation}
	\Phi(r_h)=0~.
\end{equation}
After the gauge fixing, we can directly obtain the chemical potential $\Phi_\infty$ on the AdS boundary. In order to compare with RN black holes, we rephrase the Maxwell equation
\begin{equation}
	\left(\Phi'r^{d-1}\right)'=\left(\frac{4\pi j}{f}-\frac{\chi'\Phi'}{2}\right)r^{d-1}~.
\end{equation}
Let use denote $\Phi_\text{RN}(r)$ to the gauge potential of RN black holes with same horizon and chemical potential, $i.e.$ $\Phi_\text{RN}(r)$ satisfies $\Phi_\text{RN}(r_h)=0$, $\Phi_\text{RN}(\infty)=\Phi_{\infty}$ and
\begin{equation}\label{RN}
	\left(\Phi_\text{RN}'r^{d-1}\right)'=0~.
\end{equation}
We define $\Delta\Phi=\Phi-\Phi_\text{RN}$,
\begin{equation}\label{eqphi}
	\left(\Delta\Phi' r^{d-1}\right)'=\left(\frac{4\pi j}{f}-\frac{\chi'\Phi'}{2}\right)r^{d-1}\geq0~.
\end{equation}
Since $j\geq0$, $\chi'\leq0$ and $\Phi'\leq0$. Then the "maximal principle" shows that the maximum of $\Delta\Phi$ can only attain at endpoints. Thus, we have

\begin{equation}
	\max\Delta\Phi=\Delta\Phi(r_h)=\Delta\Phi(\infty)=0~,
\end{equation}
So the relationship between the $\Phi$ and $\Phi_\text{RN}$ is
\begin{equation}
0\leq\Phi\leq\Phi_\text{RN}~.
\end{equation}
As usual, we solve function $f(r)$ in terms of $m(r)$ and $\chi(r)$
\begin{equation}
	\begin{aligned}
		&\frac{f(r){e}^{-\chi(r)}}{r^2} 
\\=2\int_{r_{h}}^{r}&\frac{\left[d \tilde{m}(x)+2\left(\Phi Q\right)-kX(x)-kx^{d-2}\right]{e}^{-\chi(x) / 2}}{x^{d+1}} \mathrm{~d} x ,
	\end{aligned}
\end{equation}
$f(r_h)'\geq0$ leads to
\begin{equation}
	 \frac{\left[d \tilde{m}(r_h)+2\Phi(r_h) Q(r_h)-kX(r_h)-kr_h^{d-2}\right]{e}^{-\chi(r_h) / 2}}{r_h^{d+1}}\geq 0,
\end{equation}
Because $X(r)\geq0$, ${e}^{-\chi(r_h) / 2}\geq0$ and $\Phi(r_h)=0$, the above inequality becomes
\begin{equation}
	\frac{d \tilde{m}(r_h)-kr_h^{d-2}}{r_h^{d+1}}\geq 0  ,
\end{equation}
Since $\tilde{m}'(r)\geq0$, we can obtain that
\begin{equation}\label{r_h}
	\begin{aligned}
		\frac{d f_{0}^{d} / 2-2(\Phi_\infty Q_\infty)-kr_h^{d-2}}{r_h^{d+1}}&=\frac{d \tilde{m}(\infty)-kr_h^{d-2}}{r_h^{d+1}}\\
&\geq\frac{d \tilde{m}(r_h)-kr_h^{d-2}}{r_h^{d+1}}\\
&\geq 0  ,
	\end{aligned}	
\end{equation}
Just like the proof in first type of generalization, we define an auxiliary function $\tilde{W}(r)$
\begin{equation}
	\begin{aligned}
	\tilde{W}(r)&=d f_{0}^{d} / 2-2(\Phi_\infty Q_\infty)+2\left(\Phi_\text{RN} Q_\infty\right)-kr^{d-2}\\&=d f_{0}^{d} / 2-2\frac{\Phi_{\infty}Q_{\infty}r_h^{d-2}}{r^{d-2}}-kr^{d-2}~.
	\end{aligned}
\end{equation}
There the $\Phi_\text{RN}$ is given by Eq.~\eqref{RN}, of which the solution reads
\begin{equation}\label{phirn}
\Phi_\text{RN}=\Phi_\infty-\frac{\Phi_\infty r_h^{d-2}}{r^{d-2}}~.
\end{equation}
One can verify that
\begin{equation}
	\tilde{W}(r_h)=d f_{0}^{d} / 2-2(\Phi_\infty Q_\infty)-kr_h^{d-2}
\end{equation}
Combing it with~\eqref{r_h}, we can obtain that
\begin{equation}
	\tilde{W}(r_h)\geq 0~.
\end{equation}
Thus the horizon $r_h$ is limited by the value of function $W(r)$. When $r\rightarrow\infty$, the inequality becomes
\begin{equation}\label{boundarycondition}
\begin{split}
	1 \leq &2\int_{r_{h}}^{\infty}[d f_{0}^{d} / 2-2(\Phi_\infty Q_\infty)+2\left(\Phi Q\right)\\
&-kx^{d-2}]\times{e}^{-\chi(x) / 2}{x^{-(d+1)}} \mathrm{~d} x\,,
\end{split}
\end{equation}
which is due to $d\tilde{m}(r)\leq d f_{0}^{d} / 2-2(\Phi_\infty Q_\infty)$ and $X(r)\geq0$. Combining it with $\Phi_\text{RN}\geq\Phi\geq0$, we can obtain
\begin{equation}
\begin{split}
	1 \leq &2\int_{r_{h}}^{\infty} [d f_{0}^{d} / 2-2(\Phi_\infty Q_\infty)+2\left(\Phi_\text{RN} Q\right)\\
&-kx^{d-2}]{e}^{-\chi(x) / 2}x^{-(d+1)} \mathrm{d}x\,.
\end{split}
\end{equation}
Because we require $j\geq0$ and the derivative of $Q(r)$ is greater than zero $Q'(r)\geq0$. Combining them with $Q\geq0$, we finally get
\begin{equation}\label{finalresult}
\begin{split}
	1 &\leq 2\int_{r_{h}}^{\infty}[d f_{0}^{d} / 2-2(\Phi_\infty Q_\infty)\\
&+2\left(\Phi_\text{RN} Q_{\infty}\right)-kx^{d-2}]{e}^{-\chi(x) / 2}x^{-(d+1)} \mathrm{~d} x\,.
\end{split}
\end{equation}
We substitute~\eqref{phirn} into the above inequality
\begin{equation}
	1 \leq 2\int_{r_{h}}^{\infty} \frac{\tilde{W}(x){e}^{-\chi(x) / 2}}{x^{d+1}} \mathrm{~d} x~,
\end{equation}
which is very similar to the proof of the first type charged generalization. The only difference is the coefficient of $1/r^{d-2}$ in the auxiliary function. So we can take the same discussion in subsection~\ref{first}
and then obtain\footnote{Like the first type of charged, if $\max W(r)\leq0$, the inequality will be broken. One can verify the mass parameter $f_{0}^{d}$ have a inequality relation: $d f_{0}^{d} / 2-2\sqrt{2\Phi_\infty Q_\infty r_h^{d-2}}\geq0$ for $k=1$ and $d f_{0}^{d} / 2\geq0$ for $k=0$.}
\begin{equation}
	1 \leq 2\int_{r_{h}}^{\infty} \frac{\tilde{W}(x){e}^{-\chi(x) / 2}}{x^{d+1}} \mathrm{~d}x\leq2{e}^{-\chi(r_2) / 2}\int_{r_{h}}^{\infty} \frac{\tilde{W}(x)}{x^{d+1}} \mathrm{~d}x\,,
\end{equation}
which yields
\begin{equation}
	 \frac{f_{0}^{d}}{r_{h}^{d}}-\frac{k}{r_{h}^{2}}-\frac{2\Phi_\infty Q_\infty}{(d-1)r_h^{d}}=2\int_{r_{h}}^{\infty} \frac{\tilde{W}(x)}{x^{d+1}} \mathrm{~d}x\geq{e}^{\chi(r_2) / 2}\geq1\,.
\end{equation}
The next step is to find the relation between $\Phi_\infty$ and $Q_\infty$. Near the infinity we have following asymptotic expansions for $\Phi$,
\begin{equation}\label{asy}
	\Phi=\Phi_\infty-\frac{Q_\infty}{(d-2)r^{d-2}}+\dots~
\end{equation}
We have already known that $\Phi\leq\Phi_\text{RN}$ for all $r\geq r_h$, then Eq.~\eqref{asy} and~\eqref{phirn} implies
\begin{equation}
	\frac{Q_\infty}{d-2}\geq\Phi_\infty r_h^{d-2}
\end{equation}
After a long journey, we finally reach the expected charged generalization
\begin{equation}
	1 \leq \frac{f_{0}^{d}}{r_{h}^{d}}-\frac{k}{r_{h}^{2}}-\frac{2\Phi_\infty Q_\infty}{(d-1)r_h^{d}}\leq \frac{f_{0}^{d}}{r_{h}^{d}}-\frac{k}{r_{h}^{2}}-\frac{2(d-2)\Phi_\infty^2}{(d-1)r_h^2}~.
\end{equation}
To saturate this inequality, we see from Eqs.~\eqref{ordinarymatter},~\eqref{boundarycondition} and~\eqref{finalresult} that $j(r)=\chi(r)=0$ and $\tilde{m}(r)=f_0^d/2$. The zero charge density and $\chi=0$ shows that $\Phi(r)=\Phi_\text{RN}(r)$. This leads to
\begin{equation}
	f(r)=1+\frac{k}{r^{2}}-\frac{f_{0}^{d}}{r^{d}}+\frac{2 Q_{\infty}^{2}}{(d-1)(d-2) r^{2 d-2}}
\end{equation}
and so the bulk geometry is a RN black hole.

\section{Penrose inequality and scheme of quantization}\label{holosec}
In above sections, we assume that the bulk geometry is asymptotically Schwarzschild AdS
so that all the quantity, especially, the total mass is defined only by bulk geometry. In
holography, when the dual field theory has nonzero external source, the total mass cannot
be read directly from the bulk metric. Instead, we have to use the so called ``holograhic
renormalization'' approach to find the total mass. In this case, our above proofs is invalid.
It is interesting to ask: can we still obtain the Penrose inequality in such case if the bulk
matters satisfy the null energy condition? In this section, we will consider the asymptotically
AdS black hole with scalar field $\phi$ as a concrete example.
\subsection{The Model}
For $(d+1)$-dimensional Einstein-scalar gravity, the theory's action reads
\begin{equation}\label{action}
S=\frac{1}{16 \pi G} \int d^{d+1} x \sqrt{-g}\left[\mathcal{R}-\frac{1}{2} \nabla_{\mu} \phi \nabla^{\mu} \phi-V(\phi)\right].
\end{equation}
Considering the static asymptotically AdS black hole with spherical/planar/hyperbolic horizon geometry, the ansatz is same as Eq.~\eqref{as}
\begin{equation} \label{ansatz}
\mathrm{d} s^{2}=-f(r) e^{-\chi(r)} \mathrm{d} t^{2}+\frac{\mathrm{d} r^{2}}{f(r)}+r^{2}\mathrm{d}\Sigma_{k, d-1}^{2}.
\end{equation}
In order to satisfy asymptotically AdS boundary condition, the function $f(r)$, $\chi(r)$  must satisfy following conditions at AdS boundary $r\rightarrow\infty$:
\begin{equation}
f(r)=r^2+ \cdots, \quad \chi(r)=\chi_{0} / r^{\alpha}+\cdots \quad \alpha>0.
\end{equation}
If $\phi(r)=0$ at any $r$, the scalar potential $V(\phi)$ will return to $-d(d-1)$ so that the theory~\eqref{action} is pure AdS gravity. Without loss of generality, assuming $\phi(r\rightarrow \infty)\rightarrow0$, we choose the potential function as
\begin{equation}
V(\phi)=-d(d-1)-\frac{1}{2} m^{2} \phi^{2}+\mathcal{O}\left(\phi^{3}\right)
\end{equation}
near the boundary.\footnote{Recall we have taken the AdS radius $\ell_{\text{AdS}}$ equal to unit in the opening of this paper.} The parameter $m$ is the mass of the scalar field. In holography, the mass-squared of the scalar field can be negative, but above the Breitenlohner-Freedman bound $m_{\text{BF}}^2$\footnote{It was first derived in Refs.~\cite{Breitenlohner:1982jf,Breitenlohner:1982bm}. Loosely speaking, the negative mass-squared below the Breitenlohner-Freedman bound $m^2<m_{\text{BF}}^2$ will lead to instability.}
\begin{equation}\label{bf}
m^2>m_{\text{BF}}^2=-\frac{d^2}{4}.
\end{equation}
According to the action \eqref{action}, equations of motion are following as:
\begin{equation}
\nabla_{\mu} \nabla^{\mu} \phi-\partial_{\phi} V=0 ,
\end{equation}
\begin{equation}
\mathcal{R}_{\mu \nu}-\frac{1}{2} \mathcal{R} g_{\mu \nu}=\frac{1}{2} \partial_{\mu} \phi \partial_{\nu} \phi+\frac{1}{2}\left(-\frac{1}{2} \nabla_{\rho} \phi \nabla^{\rho} \phi-V(\phi)\right) g_{\mu \nu} .
\end{equation}
One can check that the scalar hairy black hole solution satisfy the null energy condition. Substituting the ansatz~\eqref{ansatz} into the above equations, we can obtain
\begin{subequations} \label{eqm}
\begin{align}
\phi^{\prime \prime}+\left(\frac{f^{\prime}}{f}-\frac{\chi^{\prime}}{2}+\frac{d-1}{r}\right) \phi^{\prime}-\frac{1}{f} \partial_{\phi} V &=0, \label{eq1} \\
\frac{\chi^{\prime}}{r}+\frac{1}{d-1} \phi^{\prime 2} &=0, \label{eq2}\\
\frac{2}{r} \frac{f^{\prime}}{f}-\frac{\chi^{\prime}}{r}+\frac{2}{d-1} \frac{V}{f}+\frac{2(d-2)(f-k)}{r^{2}f·····················································} &=0. \label{eq3}
\end{align}
\end{subequations}
Near the AdS boundary, the scalar field has such asymptotic form
\begin{equation}
\phi(r)=\frac{\phi_{s}}{r^{d-\Delta}}(1+\cdots)+\frac{\phi_{v}}{r^{\Delta}}(1+\cdots)~,
\end{equation}
where $\phi_s$ and $\phi_v$ are coefficients of leading terms and $\Delta$ is the conformal dimension of the dual operator. There is the usual relationship~\cite{Witten:1998qj,Gubser:1998bc} between $\Delta$ and $m^2$
\begin{equation}
\Delta=\left(d+\sqrt{d^{2}+4 m^{2}}\right) / 2.
\end{equation}
In order to get the value of every expansion coefficient, we should expand the metric~\eqref{ansatz} at large $r$, and substitute the expansion of both scalar field $\phi$ and metric into the equations of motion~\eqref{eqm}. Then, given the boundary condition, we can solve these coefficients order by order. However, it depends on the specific form of the potential $V(\phi)$ in such solving process. With loss of generality, we consider the specific model in 4-dimensional spacetime with planar horizon geometry($k=0$) to illustrate the key feature. We take the scalar potential function~\cite{Li:2020spf} as
\begin{equation}\label{po}
V(\phi)=-6-\frac{4}{\delta^{2}} \sinh \left[\frac{\delta \phi}{2}\right]^{2}
\end{equation}
where $\delta$ is a constant. One can check the boundary's asymptotic form of potential
\begin{equation}
V(\phi)=-6-\phi^{2}+\mathcal{O}\left(\phi^{4}\right)
\end{equation}
here $\Delta=2$ and $m^2=-2$ which satisfies the Breitenlohner-Freedman bound~\eqref{bf}. Then, near the AdS boundary, we expand the metric that is determined by functions $f(r)$ and $\chi(r)$:
\begin{subequations}\label{ex}
\begin{align}
&f(r)=r^{2}\left[1+\frac{\phi_{s}^{2}}{4 r^{2}}-\frac{f_{0}^3}{r^{3}}+\mathcal{O}\left(\frac{1}{r^{4}}\right)\right], \\
&\chi(r)=\frac{\phi_{s}^{2}}{4 r^{2}}+\frac{2 \phi_{s} \phi_{v}}{3 r^{3}}+\mathcal{O}\left(\frac{1}{r^{4}}\right).
\end{align}
\end{subequations}
\subsection{Numerical check on Penrose inequality}
If we want to obtain the right holographic stress tensor $T\indices{^\mu_\nu}$ through the well-defined variational principle, the Gibbons-Hawking-York boundary term~\cite{York:1972sj,Gibbons:1976ue} should be added to the action~\eqref{action}
\begin{equation}
S_\text{GHY}=\lim _{r \rightarrow \infty} \frac{1}{8 \pi G}\int d^{3} x \sqrt{-h}K
\end{equation}
here $h_{ij}$ is the induced metric on AdS boundary, and $K$ is the trace of second fundamental form $K_{ij}$
\begin{equation}
K_{ij}=-\frac{1}{2}\mathcal{L}_nh_{ij},\qquad K=h^{ij}K_{ij},
\end{equation}
where $n^{\mu}$ is the outward pointing unit vector normal to AdS boundary. Because the action  is still divergent in AdS boundary after adding Gibbons-Hawking-York boundary term, we should introduce a boundary cosmological constant to regulate the infinity. Following standard holographic renormalization scheme~\cite{deHaro:2000vlm,Emparan:1999pm,Kraus:1999di}, the counter term for gravitational sector in this case is given by
\begin{equation}
S_{\text {c.t.}}=\lim _{r \rightarrow \infty}\frac{1}{16 \pi G}\int d^{3} x \sqrt{-h}(-4)
\end{equation}
Since mass satisfying $-\frac{d^2}{4}<m^2<1-\frac{d^2}{4}$, there are two different renormalization schemes~\cite{Klebanov:1999tb,Skenderis:2002wp,Marolf:2006nd} for scalar field $\phi(r )$ sector. For instance, if we treat $\phi_s$ as the source, we must fix the value of $\phi_s$ on the AdS boundary which is referred as standard quantization for $\phi(r )$. Then, we should add the following counter term
\begin{equation}
S_{\phi_s}=\lim _{r \rightarrow \infty}\frac{1}{16 \pi G}\int d^{3} x \sqrt{-h} \left(-\frac{1}{2}\phi^2\right).
\end{equation}
However, if we fix the value of $\phi_v$ on the boundary, the counter term we need to add is different from previous one
\begin{equation}
S_{\phi_v}=\lim _{r \rightarrow \infty}\frac{1}{16 \pi G}\int d^{3} x \sqrt{-h} \left[\phi (n^{\mu}\partial_{\mu}\phi)+\frac{1}{2}\phi^2\right].
\end{equation}
Then, we obtain the regulated action $\tilde{S}$,
\begin{equation}
\tilde{S}=S+S_\text{GHY}+S_{\text {c.t.}}+S_{\phi_{s,v}}.
\end{equation}
So far, $\tilde{S}$ is finite when $r \rightarrow \infty$. Then we can obtain the holographic stress tensor
\begin{equation}
\begin{split}
T_{\mu \nu}&=\frac{1}{16 \pi G} \lim _{r \rightarrow \infty} r[2\left(K h_{\mu \nu}-K_{\mu \nu}-2 h_{\mu \nu}\right)\\
&\left.+h_{\mu \nu}\times
\begin{cases}
\qquad-\frac{1}{2}\phi^{2}  &\\
 \phi (n^{\mu}\partial_{\mu}\phi)+\frac{1}{2}\phi^2&
\end{cases}
\right] .
\end{split}
\end{equation}
Substituting the asymptotic expansions into $T_{\mu \nu}$, we can obtain the value of $tt$ component
\begin{equation}
16 \pi GT_{tt}=
\begin{cases}
2f_{0}^3+\phi_{s}\phi_{v} &  \text{fix}~\phi_{s}\\
2f_{0}^3+2\phi_{s}\phi_{v}& \text{fix}~\phi_{v}
\end{cases}
\end{equation}
The new mass parameter $\tilde{f}_0^3$ which is defined by holographic mass/energy is relevant to the value of $T_{tt}$. In this case, $\tilde{f}_0^3$ is given by
\begin{equation}\label{fm}
\tilde{f}_0^3/2=4 \pi GT_{tt}=
\begin{cases}
f_{0}^3/2+\phi_{s}\phi_{v}/4 &  \text{fix}~\phi_{s}\\
f_{0}^3/2+\phi_{s}\phi_{v}/2& \text{fix}~\phi_{v}
\end{cases}
\end{equation}
Fixing $r_h=1$ and $\delta=1$, we can solve the equations of motion~\eqref{eqm} numerically and then read the data $\left(f_{0}^3, \phi_{s}, \phi_{v}\right) $ from the asymptotic form of Eqs.~\eqref{ex} on the boundary.
\begin{figure}[H]
	\centering
	\includegraphics[width=0.45\textwidth]{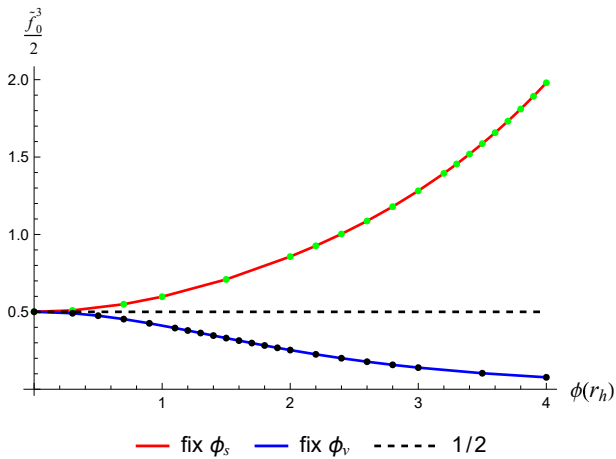}
	\caption{Mass parameter $\tilde{f}_0^3/2$ with two renormalization schemes }
	\label{holomass}
\end{figure}
In Fig.~\ref{holomass}, independent variable is the value of $\phi(r_h)$ which is the value of $\phi$ at horizon. Under the same $\phi(r_h)$, it's obvious that the value of holographic mass $M$ is different while employing two kinds of quantization schemes. In this case, the Penrose inequality is given by
\begin{equation}\label{penrose4}
\frac{4\pi M}{\Omega_{0,~2}}=\frac{\tilde{f}_0^3}{2} \geq\frac{r_h}{2}=\frac{1}{2} .
\end{equation}
The inequality is guaranteed as $\phi(r_h)$ increases if we fix $\phi_s$ on the boundary. Otherwise, the inequality will be broken  if we fix $\phi_v$ on the boundary. This means that the Penrose inequality is not generally true if we use alternative quantization in holography.

\section{Summary}
The recent holographic deduction of Penrose inequality only assumes null energy condition while the weak or dominant energy condition is required in usual geometric proof. This paper tries to make a step toward filling up  gap between these two approaches. We first discussed the AdS Penrose inequality and null energy condition from the viewpoint of pure geometry. For asymptotically Schwarzschild-AdS black hole, the matter decays fast enough so that we can read the total mass directly from asymptotically expansion of bulk metric near the AdS boundary. By the virtue of this property, we defined a "quasi-local mass"~\eqref{qlmass}  which satisfies $m(r)=M$ on the AdS boundary. What's more, due to its particular form, the derivative of $m(r)$ is non-negative~\eqref{deri} which is guaranteed by the null energy condition. Our proof indicates that the null energy condition can guarantee the Penrose inequality for black holes with planar/spherical symmetries, as expected from the holographic argument of Ref.~\cite{Engelhardt:2019btp}. The holographic argument of Ref.~\cite{Engelhardt:2019btp} also implies that null energy condition could guarantee the Penrose inequality for hyperbolically symmetric black hole, however, we find a counterexample~\eqref{ct} to shows that this is not true. These results inspire us to conjecture that the null energy condition can guarantee the Penrose inequality in asymptotically Schwarzschild AdS black hole only when the cross-section of horizon has planar or spherical topology.

Next we proposed two kinds of charged generalization for the inequality for charged black holes. The holographic argument of Ref.~\cite{Engelhardt:2019btp} implies the naive generalization~\eqref{encge}  provided null energy condition. However, counterexamples in static spherically symmetric case have been found independently in Refs.~\cite{McCormick:2019fie,Mai:2020sac} for such naive generalization. After re-examining the charge $Q$ in the inequality~\eqref{charged}, we proposed the first type of generalization which interprets $Q$ in the inequality as $Q_m$~\eqref{QM}.
However, the $Q_m$ is not defined at boundary and so cannot be interrupted as a physical quantity of dual boundary
field theory according to AdS/CFT correspondence. Thus, we proposed the second version~\eqref{second}
of charged Penrose inequality, in which the charge $Q$ is replaced by the chemical potential $\Phi_\infty$. We then gives the proofs for such two generalizations in spherically and planar symmetric cases.
Furthermore, we find that the null energy condition is not enough to guarantee the inequality in holography if the bulk geometry is asymptotically AdS but not asymptotically Schwarzschild AdS. 
In order to make this argument more explicit, this paper constructed the asymptotically AdS black holes coupled to a scalar field. Following the holographic renormalization, we find that different quantizations for $\phi(r)$ will lead to different values of holographic mass~\eqref{fm}. Then we give strong numerical evidence to show that whether the Penrose inequality holds or not will depend on quantization scheme. We note that the arguments of Refs.~\cite{Marolf:2018ldl,Engelhardt:2019btp} are regardless the topologies
of horizon and the quantization scheme. Thus, such holographic arguments would lead to
similar conclusions for different topologies and quantization schemes. However, our geometric
proofs and concrete examples show that, if we only impose null energy condition, the
whether generalized Penose inequality in asymptotically AdS spacetime is true or not will strongly depend on the topologies of horizon and the quantization schemes.
In present paper, though we propose the conjectures for general static case, we can only give the proofs in spherically and planar symmetric cases. It's worthy to examine
our conjectures in inhomogeneous cases. What's more, it is well known that in general
relativity there are many other definitions of mass, such as Kamor mass and ADM mass
and so on. In this paper, we used the holographic renormalization to define the total mass in
Penrose inequality. The question is how much the mass of different definitions influence the
structure of Penrose inequality, which is still open. Nevertheless, this paper only consider
the scalar hairy black holes. It is also interesting to consider other types of black holes,
such as vector hairy black holes.


\section*{Acknowledgements}
The work is supported by the Natural Science Foundation of China under Grant No. 12005155.

\appendix
\section{Proof of Penrose Inequality with Weak Energy Condition}\label{appendix A}
In the section \ref{nullprove}, we have considered the Penrose inequality by assuming null energy condition. Null energy condition does not require the energy density $\rho$ nonnegative but the sum of energy density and pressure density nonnegative. If matter satisfies the weak energy condition, we can find the Penrose inequality follows directly since $\rho$ is always nonnegative. Just like previous procedure, let us define a "quasi-local mass" which is known as the Hawking mass \cite{Hawking:1968qt, Szabados:2004xxa}
\begin{equation}\label{qlmweek}
	m(r)= \frac{r^{d-2}(r^2+k-f)}{2},
\end{equation}
when $r \rightarrow\infty$, one can check $m(r)$ is equal to one half of mass parameter $f_{0}^{d}$
\begin{equation}
	m(\infty)=f_{0}^{d} / 2 .
\end{equation}
Take the derivative of $m(r)$ with respect to $r$
\begin{equation}\label{m'}
	m'(r)=\frac{(d-2)k+dr^2-(d-2)f-rf'}{r^{3-d}} .
\end{equation}
Integrate the left and right sides of Eq.~\eqref{m'} from the horizon $r_h$ to $\infty$:
\begin{equation}\label{weekmass}
	m(\infty)-m(r_h)=\int_{r_{h}}^{\infty} \frac{(d-2)k+dr^2-(d-2)f-rf'}{r^{3-d}}\mathrm{~d}r .
\end{equation}
From Eq.~\eqref{e1}, the expression of $\rho$ is
\begin{equation}
	\rho=\frac{(d-1)\left[(d-2)k+dr^2-(d-2)f-rf'\right]}{2r^2}\geq 0 .
\end{equation}
Combining with weak energy condition, we can obtain
\begin{equation}
	m(\infty)-m(r_h)=f_{0}^{d} / 2-\frac{r_h^d+kr_h^{d-2}}{2} \geq 0,
\end{equation}
so that the Penrose inequality~\eqref{con1} follows. In order to see why the weak energy condition can guarantee the Penrose inequality with no difficulty, we express the integral~\eqref{weekmass} in terms of energy density $\rho$
\begin{equation}
	m(\infty)-m(r_h)=\int_{r_{h}}^{\infty} \frac{2\rho}{d-1}r^{d-1}\mathrm{~d}r .
\end{equation}
If we interpret $\rho$ as the mass density for "quasi-local mass" $m(r)$, the $m(\infty)$ must greater than or equal to $m(r_h)$ due to $\rho\geq0$. By virtue of construction of "quasi-local mass"~\eqref{qlmweek}, we can see more clearly that Penrose inequality is a stronger version of positive energy theorem for planar and spherical cases.
We conclude that the inequality is saturated only if $\chi(r)=0$ and $m(r)=m(\infty)=f_{0}^{d} / 2$, which leads to
\begin{equation}
	f(r)=r^2(1+k/r^2-f_{0}^{d}/r_h^d),~\rho=p_r=p_T=0.
\end{equation}
We can see that, if weak energy condition is satisfied, then the Penrose inequality in all three different topologies is true. The saturation appears only if the black hole is Schwarzschild-AdS black hole.
\bibliographystyle{aps}
\bibliography{rose}

\begin{thebibliography}{42}%
\makeatletter
\providecommand \@ifxundefined [1]{%
 \@ifx{#1\undefined}
}%
\providecommand \@ifnum [1]{%
 \ifnum #1\expandafter \@firstoftwo
 \else \expandafter \@secondoftwo
 \fi
}%
\providecommand \@ifx [1]{%
 \ifx #1\expandafter \@firstoftwo
 \else \expandafter \@secondoftwo
 \fi
}%
\providecommand \natexlab [1]{#1}%
\providecommand \enquote  [1]{``#1''}%
\providecommand \bibnamefont  [1]{#1}%
\providecommand \bibfnamefont [1]{#1}%
\providecommand \citenamefont [1]{#1}%
\providecommand \href@noop [0]{\@secondoftwo}%
\providecommand \href [0]{\begingroup \@sanitize@url \@href}%
\providecommand \@href[1]{\@@startlink{#1}\@@href}%
\providecommand \@@href[1]{\endgroup#1\@@endlink}%
\providecommand \@sanitize@url [0]{\catcode `\\12\catcode `\$12\catcode
  `\&12\catcode `\#12\catcode `\^12\catcode `\_12\catcode `\%12\relax}%
\providecommand \@@startlink[1]{}%
\providecommand \@@endlink[0]{}%
\providecommand \url  [0]{\begingroup\@sanitize@url \@url }%
\providecommand \@url [1]{\endgroup\@href {#1}{\urlprefix }}%
\providecommand \urlprefix  [0]{URL }%
\providecommand \Eprint [0]{\href }%
\providecommand \doibase [0]{http://dx.doi.org/}%
\providecommand \selectlanguage [0]{\@gobble}%
\providecommand \bibinfo  [0]{\@secondoftwo}%
\providecommand \bibfield  [0]{\@secondoftwo}%
\providecommand \translation [1]{[#1]}%
\providecommand \BibitemOpen [0]{}%
\providecommand \bibitemStop [0]{}%
\providecommand \bibitemNoStop [0]{.\EOS\space}%
\providecommand \EOS [0]{\spacefactor3000\relax}%
\providecommand \BibitemShut  [1]{\csname bibitem#1\endcsname}%
\let\auto@bib@innerbib\@empty
\bibitem [{\citenamefont {Bray}\ and\ \citenamefont
  {Chrusciel}(2003)}]{Bray:2003ns}%
  \BibitemOpen
  \bibfield  {author} {\bibinfo {author} {\bibfnamefont {Hubert~L.}\
  \bibnamefont {Bray}}\ and\ \bibinfo {author} {\bibfnamefont {Piotr~T.}\
  \bibnamefont {Chrusciel}},\ }\bibfield  {title} {\enquote {\bibinfo {title}
  {{The Penrose inequality}},}\ }\href@noop {} {\  (\bibinfo {year} {2003})},\
  \Eprint {http://arxiv.org/abs/gr-qc/0312047} {arXiv:gr-qc/0312047}
  \BibitemShut {NoStop}%
\bibitem [{\citenamefont {Mars}(2009)}]{Mars:2009cj}%
  \BibitemOpen
  \bibfield  {author} {\bibinfo {author} {\bibfnamefont {Marc}\ \bibnamefont
  {Mars}},\ }\bibfield  {title} {\enquote {\bibinfo {title} {{Present status of
  the Penrose inequality}},}\ }\href {\doibase 10.1088/0264-9381/26/19/193001}
  {\bibfield  {journal} {\bibinfo  {journal} {Class. Quant. Grav.}\ }\textbf
  {\bibinfo {volume} {26}},\ \bibinfo {pages} {193001} (\bibinfo {year}
  {2009})},\ \Eprint {http://arxiv.org/abs/0906.5566} {arXiv:0906.5566 [gr-qc]}
  \BibitemShut {NoStop}%
\bibitem [{\citenamefont {Schon}\ and\ \citenamefont
  {Yau}(1979)}]{Schon:1979rg}%
  \BibitemOpen
  \bibfield  {author} {\bibinfo {author} {\bibfnamefont {R.}~\bibnamefont
  {Schon}}\ and\ \bibinfo {author} {\bibfnamefont {Shing-Tung}\ \bibnamefont
  {Yau}},\ }\bibfield  {title} {\enquote {\bibinfo {title} {{On the Proof of
  the positive mass conjecture in general relativity}},}\ }\href {\doibase
  10.1007/BF01940959} {\bibfield  {journal} {\bibinfo  {journal} {Commun. Math.
  Phys.}\ }\textbf {\bibinfo {volume} {65}},\ \bibinfo {pages} {45--76}
  (\bibinfo {year} {1979})}\BibitemShut {NoStop}%
\bibitem [{\citenamefont {Schon}\ and\ \citenamefont
  {Yau}(1981)}]{Schon:1981vd}%
  \BibitemOpen
  \bibfield  {author} {\bibinfo {author} {\bibfnamefont {Richard}\ \bibnamefont
  {Schon}}\ and\ \bibinfo {author} {\bibfnamefont {Shing-Tung}\ \bibnamefont
  {Yau}},\ }\bibfield  {title} {\enquote {\bibinfo {title} {{Proof of the
  positive mass theorem. 2.}}}\ }\href {\doibase 10.1007/BF01942062} {\bibfield
   {journal} {\bibinfo  {journal} {Commun. Math. Phys.}\ }\textbf {\bibinfo
  {volume} {79}},\ \bibinfo {pages} {231--260} (\bibinfo {year}
  {1981})}\BibitemShut {NoStop}%
\bibitem [{\citenamefont {Bardeen}\ \emph {et~al.}(1973)\citenamefont
  {Bardeen}, \citenamefont {Carter},\ and\ \citenamefont
  {Hawking}}]{Bardeen:1973gs}%
  \BibitemOpen
  \bibfield  {author} {\bibinfo {author} {\bibfnamefont {James~M.}\
  \bibnamefont {Bardeen}}, \bibinfo {author} {\bibfnamefont {B.}~\bibnamefont
  {Carter}}, \ and\ \bibinfo {author} {\bibfnamefont {S.~W.}\ \bibnamefont
  {Hawking}},\ }\bibfield  {title} {\enquote {\bibinfo {title} {{The Four laws
  of black hole mechanics}},}\ }\href {\doibase 10.1007/BF01645742} {\bibfield
  {journal} {\bibinfo  {journal} {Commun. Math. Phys.}\ }\textbf {\bibinfo
  {volume} {31}},\ \bibinfo {pages} {161--170} (\bibinfo {year}
  {1973})}\BibitemShut {NoStop}%
\bibitem [{\citenamefont {Wald}(2001)}]{Wald:1999vt}%
  \BibitemOpen
  \bibfield  {author} {\bibinfo {author} {\bibfnamefont {Robert~M.}\
  \bibnamefont {Wald}},\ }\bibfield  {title} {\enquote {\bibinfo {title} {{The
  thermodynamics of black holes}},}\ }\href {\doibase 10.12942/lrr-2001-6}
  {\bibfield  {journal} {\bibinfo  {journal} {Living Rev. Rel.}\ }\textbf
  {\bibinfo {volume} {4}},\ \bibinfo {pages} {6} (\bibinfo {year} {2001})},\
  \Eprint {http://arxiv.org/abs/gr-qc/9912119} {arXiv:gr-qc/9912119}
  \BibitemShut {NoStop}%
\bibitem [{\citenamefont {Ben-Dov}(2004)}]{Ben-Dov:2004lmn}%
  \BibitemOpen
  \bibfield  {author} {\bibinfo {author} {\bibfnamefont {Ishai}\ \bibnamefont
  {Ben-Dov}},\ }\bibfield  {title} {\enquote {\bibinfo {title} {{The Penrose
  inequality and apparent horizons}},}\ }\href {\doibase
  10.1103/PhysRevD.70.124031} {\bibfield  {journal} {\bibinfo  {journal} {Phys.
  Rev. D}\ }\textbf {\bibinfo {volume} {70}},\ \bibinfo {pages} {124031}
  (\bibinfo {year} {2004})},\ \Eprint {http://arxiv.org/abs/gr-qc/0408066}
  {arXiv:gr-qc/0408066} \BibitemShut {NoStop}%
\bibitem [{\citenamefont {Penrose}(1973)}]{Penrose:1973um}%
  \BibitemOpen
  \bibfield  {author} {\bibinfo {author} {\bibfnamefont {R.}~\bibnamefont
  {Penrose}},\ }\bibfield  {title} {\enquote {\bibinfo {title} {{Naked
  singularities}},}\ }\href {\doibase 10.1111/j.1749-6632.1973.tb41447.x}
  {\bibfield  {journal} {\bibinfo  {journal} {Annals N. Y. Acad. Sci.}\
  }\textbf {\bibinfo {volume} {224}},\ \bibinfo {pages} {125--134} (\bibinfo
  {year} {1973})}\BibitemShut {NoStop}%
\bibitem [{\citenamefont {Bray}(2001)}]{10.4310/jdg/1090349428}%
  \BibitemOpen
  \bibfield  {author} {\bibinfo {author} {\bibfnamefont {Hubert~L.}\
  \bibnamefont {Bray}},\ }\bibfield  {title} {\enquote {\bibinfo {title}
  {{Proof of the Riemannian Penrose Inequality Using the Positive Mass
  Theorem}},}\ }\href {\doibase 10.4310/jdg/1090349428} {\bibfield  {journal}
  {\bibinfo  {journal} {Journal of Differential Geometry}\ }\textbf {\bibinfo
  {volume} {59}},\ \bibinfo {pages} {177 -- 267} (\bibinfo {year}
  {2001})}\BibitemShut {NoStop}%
\bibitem [{\citenamefont {Huisken}\ and\ \citenamefont
  {Ilmanen}(2001)}]{10.4310/jdg/1090349447}%
  \BibitemOpen
  \bibfield  {author} {\bibinfo {author} {\bibfnamefont {Gerhard}\ \bibnamefont
  {Huisken}}\ and\ \bibinfo {author} {\bibfnamefont {Tom}\ \bibnamefont
  {Ilmanen}},\ }\bibfield  {title} {\enquote {\bibinfo {title} {{The Inverse
  Mean Curvature Flow and the Riemannian Penrose Inequality}},}\ }\href
  {\doibase 10.4310/jdg/1090349447} {\bibfield  {journal} {\bibinfo  {journal}
  {Journal of Differential Geometry}\ }\textbf {\bibinfo {volume} {59}},\
  \bibinfo {pages} {353 -- 437} (\bibinfo {year} {2001})}\BibitemShut {NoStop}%
\bibitem [{\citenamefont {Itkin}\ and\ \citenamefont
  {Oz}(2012)}]{Itkin:2011ph}%
  \BibitemOpen
  \bibfield  {author} {\bibinfo {author} {\bibfnamefont {Igor}\ \bibnamefont
  {Itkin}}\ and\ \bibinfo {author} {\bibfnamefont {Yaron}\ \bibnamefont {Oz}},\
  }\bibfield  {title} {\enquote {\bibinfo {title} {{Penrose Inequality for
  Asymptotically AdS Spaces}},}\ }\href {\doibase
  10.1016/j.physletb.2012.01.007} {\bibfield  {journal} {\bibinfo  {journal}
  {Phys. Lett. B}\ }\textbf {\bibinfo {volume} {708}},\ \bibinfo {pages}
  {307--308} (\bibinfo {year} {2012})},\ \Eprint
  {http://arxiv.org/abs/1106.2683} {arXiv:1106.2683 [hep-th]} \BibitemShut
  {NoStop}%
\bibitem [{\citenamefont {Emparan}\ \emph {et~al.}(1999)\citenamefont
  {Emparan}, \citenamefont {Johnson},\ and\ \citenamefont
  {Myers}}]{Emparan:1999pm}%
  \BibitemOpen
  \bibfield  {author} {\bibinfo {author} {\bibfnamefont {Roberto}\ \bibnamefont
  {Emparan}}, \bibinfo {author} {\bibfnamefont {Clifford~V.}\ \bibnamefont
  {Johnson}}, \ and\ \bibinfo {author} {\bibfnamefont {Robert~C.}\ \bibnamefont
  {Myers}},\ }\bibfield  {title} {\enquote {\bibinfo {title} {{Surface terms as
  counterterms in the AdS / CFT correspondence}},}\ }\href {\doibase
  10.1103/PhysRevD.60.104001} {\bibfield  {journal} {\bibinfo  {journal} {Phys.
  Rev. D}\ }\textbf {\bibinfo {volume} {60}},\ \bibinfo {pages} {104001}
  (\bibinfo {year} {1999})},\ \Eprint {http://arxiv.org/abs/hep-th/9903238}
  {arXiv:hep-th/9903238} \BibitemShut {NoStop}%
\bibitem [{\citenamefont {Witten}(1998)}]{Witten:1998qj}%
  \BibitemOpen
  \bibfield  {author} {\bibinfo {author} {\bibfnamefont {Edward}\ \bibnamefont
  {Witten}},\ }\bibfield  {title} {\enquote {\bibinfo {title} {{Anti-de Sitter
  space and holography}},}\ }\href {\doibase 10.4310/ATMP.1998.v2.n2.a2}
  {\bibfield  {journal} {\bibinfo  {journal} {Adv. Theor. Math. Phys.}\
  }\textbf {\bibinfo {volume} {2}},\ \bibinfo {pages} {253--291} (\bibinfo
  {year} {1998})},\ \Eprint {http://arxiv.org/abs/hep-th/9802150}
  {arXiv:hep-th/9802150} \BibitemShut {NoStop}%
\bibitem [{\citenamefont {Aharony}\ \emph {et~al.}(2000)\citenamefont
  {Aharony}, \citenamefont {Gubser}, \citenamefont {Maldacena}, \citenamefont
  {Ooguri},\ and\ \citenamefont {Oz}}]{Aharony:1999ti}%
  \BibitemOpen
  \bibfield  {author} {\bibinfo {author} {\bibfnamefont {Ofer}\ \bibnamefont
  {Aharony}}, \bibinfo {author} {\bibfnamefont {Steven~S.}\ \bibnamefont
  {Gubser}}, \bibinfo {author} {\bibfnamefont {Juan~Martin}\ \bibnamefont
  {Maldacena}}, \bibinfo {author} {\bibfnamefont {Hirosi}\ \bibnamefont
  {Ooguri}}, \ and\ \bibinfo {author} {\bibfnamefont {Yaron}\ \bibnamefont
  {Oz}},\ }\bibfield  {title} {\enquote {\bibinfo {title} {{Large N field
  theories, string theory and gravity}},}\ }\href {\doibase
  10.1016/S0370-1573(99)00083-6} {\bibfield  {journal} {\bibinfo  {journal}
  {Phys. Rept.}\ }\textbf {\bibinfo {volume} {323}},\ \bibinfo {pages}
  {183--386} (\bibinfo {year} {2000})},\ \Eprint
  {http://arxiv.org/abs/hep-th/9905111} {arXiv:hep-th/9905111} \BibitemShut
  {NoStop}%
\bibitem [{\citenamefont {Ryu}\ and\ \citenamefont
  {Takayanagi}(2006)}]{Ryu:2006bv}%
  \BibitemOpen
  \bibfield  {author} {\bibinfo {author} {\bibfnamefont {Shinsei}\ \bibnamefont
  {Ryu}}\ and\ \bibinfo {author} {\bibfnamefont {Tadashi}\ \bibnamefont
  {Takayanagi}},\ }\bibfield  {title} {\enquote {\bibinfo {title} {{Holographic
  derivation of entanglement entropy from AdS/CFT}},}\ }\href {\doibase
  10.1103/PhysRevLett.96.181602} {\bibfield  {journal} {\bibinfo  {journal}
  {Phys. Rev. Lett.}\ }\textbf {\bibinfo {volume} {96}},\ \bibinfo {pages}
  {181602} (\bibinfo {year} {2006})},\ \Eprint
  {http://arxiv.org/abs/hep-th/0603001} {arXiv:hep-th/0603001} \BibitemShut
  {NoStop}%
\bibitem [{\citenamefont {Engelhardt}\ and\ \citenamefont
  {Wall}(2018)}]{Engelhardt:2017aux}%
  \BibitemOpen
  \bibfield  {author} {\bibinfo {author} {\bibfnamefont {Netta}\ \bibnamefont
  {Engelhardt}}\ and\ \bibinfo {author} {\bibfnamefont {Aron~C.}\ \bibnamefont
  {Wall}},\ }\bibfield  {title} {\enquote {\bibinfo {title} {{Decoding the
  Apparent Horizon: Coarse-Grained Holographic Entropy}},}\ }\href {\doibase
  10.1103/PhysRevLett.121.211301} {\bibfield  {journal} {\bibinfo  {journal}
  {Phys. Rev. Lett.}\ }\textbf {\bibinfo {volume} {121}},\ \bibinfo {pages}
  {211301} (\bibinfo {year} {2018})},\ \Eprint
  {http://arxiv.org/abs/1706.02038} {arXiv:1706.02038 [hep-th]} \BibitemShut
  {NoStop}%
\bibitem [{\citenamefont {Marolf}(2018)}]{Marolf:2018ldl}%
  \BibitemOpen
  \bibfield  {author} {\bibinfo {author} {\bibfnamefont {Donald}\ \bibnamefont
  {Marolf}},\ }\bibfield  {title} {\enquote {\bibinfo {title} {{Microcanonical
  Path Integrals and the Holography of small Black Hole Interiors}},}\ }\href
  {\doibase 10.1007/JHEP09(2018)114} {\bibfield  {journal} {\bibinfo  {journal}
  {JHEP}\ }\textbf {\bibinfo {volume} {09}},\ \bibinfo {pages} {114} (\bibinfo
  {year} {2018})},\ \Eprint {http://arxiv.org/abs/1808.00394} {arXiv:1808.00394
  [hep-th]} \BibitemShut {NoStop}%
\bibitem [{\citenamefont {Engelhardt}\ and\ \citenamefont
  {Horowitz}(2019)}]{Engelhardt:2019btp}%
  \BibitemOpen
  \bibfield  {author} {\bibinfo {author} {\bibfnamefont {Netta}\ \bibnamefont
  {Engelhardt}}\ and\ \bibinfo {author} {\bibfnamefont {Gary~T.}\ \bibnamefont
  {Horowitz}},\ }\bibfield  {title} {\enquote {\bibinfo {title} {{Holographic
  argument for the Penrose inequality in AdS spacetimes}},}\ }\href {\doibase
  10.1103/PhysRevD.99.126009} {\bibfield  {journal} {\bibinfo  {journal} {Phys.
  Rev. D}\ }\textbf {\bibinfo {volume} {99}},\ \bibinfo {pages} {126009}
  (\bibinfo {year} {2019})},\ \Eprint {http://arxiv.org/abs/1903.00555}
  {arXiv:1903.00555 [hep-th]} \BibitemShut {NoStop}%
\bibitem [{\citenamefont {McCormick}(2020)}]{McCormick:2019fie}%
  \BibitemOpen
  \bibfield  {author} {\bibinfo {author} {\bibfnamefont {Stephen}\ \bibnamefont
  {McCormick}},\ }\bibfield  {title} {\enquote {\bibinfo {title} {{On the
  charged Riemannian Penrose inequality with charged matter}},}\ }\href
  {\doibase 10.1088/1361-6382/ab50a8} {\bibfield  {journal} {\bibinfo
  {journal} {Class. Quant. Grav.}\ }\textbf {\bibinfo {volume} {37}},\ \bibinfo
  {pages} {015007} (\bibinfo {year} {2020})},\ \Eprint
  {http://arxiv.org/abs/1907.07967} {arXiv:1907.07967 [gr-qc]} \BibitemShut
  {NoStop}%
\bibitem [{\citenamefont {Mai}\ and\ \citenamefont {Yang}(2021)}]{Mai:2020sac}%
  \BibitemOpen
  \bibfield  {author} {\bibinfo {author} {\bibfnamefont {Zhan-Feng}\
  \bibnamefont {Mai}}\ and\ \bibinfo {author} {\bibfnamefont {Run-Qiu}\
  \bibnamefont {Yang}},\ }\bibfield  {title} {\enquote {\bibinfo {title}
  {{Stability analysis of a charged black hole with a nonlinear complex scalar
  field}},}\ }\href {\doibase 10.1103/PhysRevD.104.044008} {\bibfield
  {journal} {\bibinfo  {journal} {Phys. Rev. D}\ }\textbf {\bibinfo {volume}
  {104}},\ \bibinfo {pages} {044008} (\bibinfo {year} {2021})},\ \Eprint
  {http://arxiv.org/abs/2101.00026} {arXiv:2101.00026 [gr-qc]} \BibitemShut
  {NoStop}%
\bibitem [{\citenamefont {Lee}\ and\ \citenamefont
  {Neves}(2015)}]{Lee:2015xha}%
  \BibitemOpen
  \bibfield  {author} {\bibinfo {author} {\bibfnamefont {Dan~A.}\ \bibnamefont
  {Lee}}\ and\ \bibinfo {author} {\bibfnamefont {Andr\'e}\ \bibnamefont
  {Neves}},\ }\bibfield  {title} {\enquote {\bibinfo {title} {{The Penrose
  Inequality for Asymptotically Locally Hyperbolic Spaces with Nonpositive
  Mass}},}\ }\href {\doibase 10.1007/s00220-015-2421-x} {\bibfield  {journal}
  {\bibinfo  {journal} {Commun. Math. Phys.}\ }\textbf {\bibinfo {volume}
  {339}},\ \bibinfo {pages} {327--352} (\bibinfo {year} {2015})}\BibitemShut
  {NoStop}%
\bibitem [{\citenamefont {Husain}\ and\ \citenamefont
  {Singh}(2017)}]{Husain:2017cmj}%
  \BibitemOpen
  \bibfield  {author} {\bibinfo {author} {\bibfnamefont {Viqar}\ \bibnamefont
  {Husain}}\ and\ \bibinfo {author} {\bibfnamefont {Suprit}\ \bibnamefont
  {Singh}},\ }\bibfield  {title} {\enquote {\bibinfo {title} {{Penrose
  inequality in anti\textendash{}de Sitter space}},}\ }\href {\doibase
  10.1103/PhysRevD.96.104055} {\bibfield  {journal} {\bibinfo  {journal} {Phys.
  Rev. D}\ }\textbf {\bibinfo {volume} {96}},\ \bibinfo {pages} {104055}
  (\bibinfo {year} {2017})},\ \Eprint {http://arxiv.org/abs/1709.02395}
  {arXiv:1709.02395 [gr-qc]} \BibitemShut {NoStop}%
\bibitem [{\citenamefont {Shi}\ and\ \citenamefont
  {Zhu}(2021)}]{shi2021regularity}%
  \BibitemOpen
  \bibfield  {author} {\bibinfo {author} {\bibfnamefont {Yuguang}\ \bibnamefont
  {Shi}}\ and\ \bibinfo {author} {\bibfnamefont {Jintian}\ \bibnamefont
  {Zhu}},\ }\href@noop {} {\enquote {\bibinfo {title} {Regularity of inverse
  mean curvature flow in asymptotically hyperbolic manifolds with dimension
  $3$},}\ } (\bibinfo {year} {2021}),\ \Eprint
  {http://arxiv.org/abs/1811.06158} {arXiv:1811.06158 [math.DG]} \BibitemShut
  {NoStop}%
\bibitem [{\citenamefont {Ashtekar}\ and\ \citenamefont
  {Das}(2000)}]{Ashtekar:1999jx}%
  \BibitemOpen
  \bibfield  {author} {\bibinfo {author} {\bibfnamefont {Abhay}\ \bibnamefont
  {Ashtekar}}\ and\ \bibinfo {author} {\bibfnamefont {Saurya}\ \bibnamefont
  {Das}},\ }\bibfield  {title} {\enquote {\bibinfo {title} {{Asymptotically
  Anti-de Sitter space-times: Conserved quantities}},}\ }\href {\doibase
  10.1088/0264-9381/17/2/101} {\bibfield  {journal} {\bibinfo  {journal}
  {Class. Quant. Grav.}\ }\textbf {\bibinfo {volume} {17}},\ \bibinfo {pages}
  {L17--L30} (\bibinfo {year} {2000})},\ \Eprint
  {http://arxiv.org/abs/hep-th/9911230} {arXiv:hep-th/9911230} \BibitemShut
  {NoStop}%
\bibitem [{\citenamefont {Balasubramanian}\ and\ \citenamefont
  {Kraus}(1999)}]{Balasubramanian:1999re}%
  \BibitemOpen
  \bibfield  {author} {\bibinfo {author} {\bibfnamefont {Vijay}\ \bibnamefont
  {Balasubramanian}}\ and\ \bibinfo {author} {\bibfnamefont {Per}\ \bibnamefont
  {Kraus}},\ }\bibfield  {title} {\enquote {\bibinfo {title} {{A Stress tensor
  for Anti-de Sitter gravity}},}\ }\href {\doibase 10.1007/s002200050764}
  {\bibfield  {journal} {\bibinfo  {journal} {Commun. Math. Phys.}\ }\textbf
  {\bibinfo {volume} {208}},\ \bibinfo {pages} {413--428} (\bibinfo {year}
  {1999})},\ \Eprint {http://arxiv.org/abs/hep-th/9902121}
  {arXiv:hep-th/9902121} \BibitemShut {NoStop}%
\bibitem [{\citenamefont {Myers}(1999)}]{Myers:1999psa}%
  \BibitemOpen
  \bibfield  {author} {\bibinfo {author} {\bibfnamefont {Robert~C.}\
  \bibnamefont {Myers}},\ }\bibfield  {title} {\enquote {\bibinfo {title}
  {{Stress tensors and Casimir energies in the AdS / CFT correspondence}},}\
  }\href {\doibase 10.1103/PhysRevD.60.046002} {\bibfield  {journal} {\bibinfo
  {journal} {Phys. Rev. D}\ }\textbf {\bibinfo {volume} {60}},\ \bibinfo
  {pages} {046002} (\bibinfo {year} {1999})},\ \Eprint
  {http://arxiv.org/abs/hep-th/9903203} {arXiv:hep-th/9903203} \BibitemShut
  {NoStop}%
\bibitem [{\citenamefont {de~Haro}\ \emph {et~al.}(2001)\citenamefont
  {de~Haro}, \citenamefont {Solodukhin},\ and\ \citenamefont
  {Skenderis}}]{deHaro:2000vlm}%
  \BibitemOpen
  \bibfield  {author} {\bibinfo {author} {\bibfnamefont {Sebastian}\
  \bibnamefont {de~Haro}}, \bibinfo {author} {\bibfnamefont {Sergey~N.}\
  \bibnamefont {Solodukhin}}, \ and\ \bibinfo {author} {\bibfnamefont {Kostas}\
  \bibnamefont {Skenderis}},\ }\bibfield  {title} {\enquote {\bibinfo {title}
  {{Holographic reconstruction of space-time and renormalization in the AdS /
  CFT correspondence}},}\ }\href {\doibase 10.1007/s002200100381} {\bibfield
  {journal} {\bibinfo  {journal} {Commun. Math. Phys.}\ }\textbf {\bibinfo
  {volume} {217}},\ \bibinfo {pages} {595--622} (\bibinfo {year} {2001})},\
  \Eprint {http://arxiv.org/abs/hep-th/0002230} {arXiv:hep-th/0002230}
  \BibitemShut {NoStop}%
\bibitem [{\citenamefont {Skenderis}(2002)}]{Skenderis:2002wp}%
  \BibitemOpen
  \bibfield  {author} {\bibinfo {author} {\bibfnamefont {Kostas}\ \bibnamefont
  {Skenderis}},\ }\bibfield  {title} {\enquote {\bibinfo {title} {{Lecture
  notes on holographic renormalization}},}\ }\href {\doibase
  10.1088/0264-9381/19/22/306} {\bibfield  {journal} {\bibinfo  {journal}
  {Class. Quant. Grav.}\ }\textbf {\bibinfo {volume} {19}},\ \bibinfo {pages}
  {5849--5876} (\bibinfo {year} {2002})},\ \Eprint
  {http://arxiv.org/abs/hep-th/0209067} {arXiv:hep-th/0209067} \BibitemShut
  {NoStop}%
\bibitem [{\citenamefont {Hawking}(1972)}]{Hawking:1971vc}%
  \BibitemOpen
  \bibfield  {author} {\bibinfo {author} {\bibfnamefont {S.~W.}\ \bibnamefont
  {Hawking}},\ }\bibfield  {title} {\enquote {\bibinfo {title} {{Black holes in
  general relativity}},}\ }\href {\doibase 10.1007/BF01877517} {\bibfield
  {journal} {\bibinfo  {journal} {Commun. Math. Phys.}\ }\textbf {\bibinfo
  {volume} {25}},\ \bibinfo {pages} {152--166} (\bibinfo {year}
  {1972})}\BibitemShut {NoStop}%
\bibitem [{\citenamefont {Hawking}\ and\ \citenamefont
  {Ellis}(2011)}]{Hawking:1973uf}%
  \BibitemOpen
  \bibfield  {author} {\bibinfo {author} {\bibfnamefont {S.~W.}\ \bibnamefont
  {Hawking}}\ and\ \bibinfo {author} {\bibfnamefont {G.~F.~R.}\ \bibnamefont
  {Ellis}},\ }\href {\doibase 10.1017/CBO9780511524646} {\emph {\bibinfo
  {title} {{The Large Scale Structure of Space-Time}}}},\ Cambridge Monographs
  on Mathematical Physics\ (\bibinfo  {publisher} {Cambridge University
  Press},\ \bibinfo {year} {2011})\BibitemShut {NoStop}%
\bibitem [{\citenamefont {Chru\'sciel}\ \emph {et~al.}(2021)\citenamefont
  {Chru\'sciel}, \citenamefont {Delay},\ and\ \citenamefont
  {Wutte}}]{Chrusciel:2021ufc}%
  \BibitemOpen
  \bibfield  {author} {\bibinfo {author} {\bibfnamefont {Piotr~T.}\
  \bibnamefont {Chru\'sciel}}, \bibinfo {author} {\bibfnamefont {Erwann}\
  \bibnamefont {Delay}}, \ and\ \bibinfo {author} {\bibfnamefont {Raphaela}\
  \bibnamefont {Wutte}},\ }\bibfield  {title} {\enquote {\bibinfo {title}
  {{Hyperbolic energy and Maskit gluings}},}\ }\href@noop {} {\  (\bibinfo
  {year} {2021})},\ \Eprint {http://arxiv.org/abs/2112.00095} {arXiv:2112.00095
  [math.DG]} \BibitemShut {NoStop}%
\bibitem [{\citenamefont {Breitenlohner}\ and\ \citenamefont
  {Freedman}(1982{\natexlab{a}})}]{Breitenlohner:1982jf}%
  \BibitemOpen
  \bibfield  {author} {\bibinfo {author} {\bibfnamefont {Peter}\ \bibnamefont
  {Breitenlohner}}\ and\ \bibinfo {author} {\bibfnamefont {Daniel~Z.}\
  \bibnamefont {Freedman}},\ }\bibfield  {title} {\enquote {\bibinfo {title}
  {{Stability in Gauged Extended Supergravity}},}\ }\href {\doibase
  10.1016/0003-4916(82)90116-6} {\bibfield  {journal} {\bibinfo  {journal}
  {Annals Phys.}\ }\textbf {\bibinfo {volume} {144}},\ \bibinfo {pages} {249}
  (\bibinfo {year} {1982}{\natexlab{a}})}\BibitemShut {NoStop}%
\bibitem [{\citenamefont {Breitenlohner}\ and\ \citenamefont
  {Freedman}(1982{\natexlab{b}})}]{Breitenlohner:1982bm}%
  \BibitemOpen
  \bibfield  {author} {\bibinfo {author} {\bibfnamefont {Peter}\ \bibnamefont
  {Breitenlohner}}\ and\ \bibinfo {author} {\bibfnamefont {Daniel~Z.}\
  \bibnamefont {Freedman}},\ }\bibfield  {title} {\enquote {\bibinfo {title}
  {{Positive Energy in anti-De Sitter Backgrounds and Gauged Extended
  Supergravity}},}\ }\href {\doibase 10.1016/0370-2693(82)90643-8} {\bibfield
  {journal} {\bibinfo  {journal} {Phys. Lett. B}\ }\textbf {\bibinfo {volume}
  {115}},\ \bibinfo {pages} {197--201} (\bibinfo {year}
  {1982}{\natexlab{b}})}\BibitemShut {NoStop}%
\bibitem [{\citenamefont {Gubser}\ \emph {et~al.}(1998)\citenamefont {Gubser},
  \citenamefont {Klebanov},\ and\ \citenamefont {Polyakov}}]{Gubser:1998bc}%
  \BibitemOpen
  \bibfield  {author} {\bibinfo {author} {\bibfnamefont {S.~S.}\ \bibnamefont
  {Gubser}}, \bibinfo {author} {\bibfnamefont {Igor~R.}\ \bibnamefont
  {Klebanov}}, \ and\ \bibinfo {author} {\bibfnamefont {Alexander~M.}\
  \bibnamefont {Polyakov}},\ }\bibfield  {title} {\enquote {\bibinfo {title}
  {{Gauge theory correlators from noncritical string theory}},}\ }\href
  {\doibase 10.1016/S0370-2693(98)00377-3} {\bibfield  {journal} {\bibinfo
  {journal} {Phys. Lett. B}\ }\textbf {\bibinfo {volume} {428}},\ \bibinfo
  {pages} {105--114} (\bibinfo {year} {1998})},\ \Eprint
  {http://arxiv.org/abs/hep-th/9802109} {arXiv:hep-th/9802109} \BibitemShut
  {NoStop}%
\bibitem [{\citenamefont {Li}(2021)}]{Li:2020spf}%
  \BibitemOpen
  \bibfield  {author} {\bibinfo {author} {\bibfnamefont {Li}~\bibnamefont
  {Li}},\ }\bibfield  {title} {\enquote {\bibinfo {title} {{On Thermodynamics
  of AdS Black Holes with Scalar Hair}},}\ }\href {\doibase
  10.1016/j.physletb.2021.136123} {\bibfield  {journal} {\bibinfo  {journal}
  {Phys. Lett. B}\ }\textbf {\bibinfo {volume} {815}},\ \bibinfo {pages}
  {136123} (\bibinfo {year} {2021})},\ \Eprint
  {http://arxiv.org/abs/2008.05597} {arXiv:2008.05597 [gr-qc]} \BibitemShut
  {NoStop}%
\bibitem [{\citenamefont {York}(1972)}]{York:1972sj}%
  \BibitemOpen
  \bibfield  {author} {\bibinfo {author} {\bibfnamefont {James~W.}\
  \bibnamefont {York}, \bibfnamefont {Jr.}},\ }\bibfield  {title} {\enquote
  {\bibinfo {title} {{Role of conformal three geometry in the dynamics of
  gravitation}},}\ }\href {\doibase 10.1103/PhysRevLett.28.1082} {\bibfield
  {journal} {\bibinfo  {journal} {Phys. Rev. Lett.}\ }\textbf {\bibinfo
  {volume} {28}},\ \bibinfo {pages} {1082--1085} (\bibinfo {year}
  {1972})}\BibitemShut {NoStop}%
\bibitem [{\citenamefont {Gibbons}\ and\ \citenamefont
  {Hawking}(1977)}]{Gibbons:1976ue}%
  \BibitemOpen
  \bibfield  {author} {\bibinfo {author} {\bibfnamefont {G.~W.}\ \bibnamefont
  {Gibbons}}\ and\ \bibinfo {author} {\bibfnamefont {S.~W.}\ \bibnamefont
  {Hawking}},\ }\bibfield  {title} {\enquote {\bibinfo {title} {{Action
  Integrals and Partition Functions in Quantum Gravity}},}\ }\href {\doibase
  10.1103/PhysRevD.15.2752} {\bibfield  {journal} {\bibinfo  {journal} {Phys.
  Rev. D}\ }\textbf {\bibinfo {volume} {15}},\ \bibinfo {pages} {2752--2756}
  (\bibinfo {year} {1977})}\BibitemShut {NoStop}%
\bibitem [{\citenamefont {Kraus}\ \emph {et~al.}(1999)\citenamefont {Kraus},
  \citenamefont {Larsen},\ and\ \citenamefont {Siebelink}}]{Kraus:1999di}%
  \BibitemOpen
  \bibfield  {author} {\bibinfo {author} {\bibfnamefont {Per}\ \bibnamefont
  {Kraus}}, \bibinfo {author} {\bibfnamefont {Finn}\ \bibnamefont {Larsen}}, \
  and\ \bibinfo {author} {\bibfnamefont {Ruud}\ \bibnamefont {Siebelink}},\
  }\bibfield  {title} {\enquote {\bibinfo {title} {{The gravitational action in
  asymptotically AdS and flat space-times}},}\ }\href {\doibase
  10.1016/S0550-3213(99)00549-0} {\bibfield  {journal} {\bibinfo  {journal}
  {Nucl. Phys. B}\ }\textbf {\bibinfo {volume} {563}},\ \bibinfo {pages}
  {259--278} (\bibinfo {year} {1999})},\ \Eprint
  {http://arxiv.org/abs/hep-th/9906127} {arXiv:hep-th/9906127} \BibitemShut
  {NoStop}%
\bibitem [{\citenamefont {Klebanov}\ and\ \citenamefont
  {Witten}(1999)}]{Klebanov:1999tb}%
  \BibitemOpen
  \bibfield  {author} {\bibinfo {author} {\bibfnamefont {Igor~R.}\ \bibnamefont
  {Klebanov}}\ and\ \bibinfo {author} {\bibfnamefont {Edward}\ \bibnamefont
  {Witten}},\ }\bibfield  {title} {\enquote {\bibinfo {title} {{AdS / CFT
  correspondence and symmetry breaking}},}\ }\href {\doibase
  10.1016/S0550-3213(99)00387-9} {\bibfield  {journal} {\bibinfo  {journal}
  {Nucl. Phys. B}\ }\textbf {\bibinfo {volume} {556}},\ \bibinfo {pages}
  {89--114} (\bibinfo {year} {1999})},\ \Eprint
  {http://arxiv.org/abs/hep-th/9905104} {arXiv:hep-th/9905104} \BibitemShut
  {NoStop}%
\bibitem [{\citenamefont {Marolf}\ and\ \citenamefont
  {Ross}(2006)}]{Marolf:2006nd}%
  \BibitemOpen
  \bibfield  {author} {\bibinfo {author} {\bibfnamefont {Donald}\ \bibnamefont
  {Marolf}}\ and\ \bibinfo {author} {\bibfnamefont {Simon~F.}\ \bibnamefont
  {Ross}},\ }\bibfield  {title} {\enquote {\bibinfo {title} {{Boundary
  Conditions and New Dualities: Vector Fields in AdS/CFT}},}\ }\href {\doibase
  10.1088/1126-6708/2006/11/085} {\bibfield  {journal} {\bibinfo  {journal}
  {JHEP}\ }\textbf {\bibinfo {volume} {11}},\ \bibinfo {pages} {085} (\bibinfo
  {year} {2006})},\ \Eprint {http://arxiv.org/abs/hep-th/0606113}
  {arXiv:hep-th/0606113} \BibitemShut {NoStop}%
\bibitem [{\citenamefont {Hawking}(1968)}]{Hawking:1968qt}%
  \BibitemOpen
  \bibfield  {author} {\bibinfo {author} {\bibfnamefont {Stephen}\ \bibnamefont
  {Hawking}},\ }\bibfield  {title} {\enquote {\bibinfo {title} {{Gravitational
  radiation in an expanding universe}},}\ }\href {\doibase 10.1063/1.1664615}
  {\bibfield  {journal} {\bibinfo  {journal} {J. Math. Phys.}\ }\textbf
  {\bibinfo {volume} {9}},\ \bibinfo {pages} {598--604} (\bibinfo {year}
  {1968})}\BibitemShut {NoStop}%
\bibitem [{\citenamefont {Szabados}(2004)}]{Szabados:2004xxa}%
  \BibitemOpen
  \bibfield  {author} {\bibinfo {author} {\bibfnamefont {Laszlo~B.}\
  \bibnamefont {Szabados}},\ }\bibfield  {title} {\enquote {\bibinfo {title}
  {{Quasi-Local Energy-Momentum and Angular Momentum in GR: A Review
  Article}},}\ }\href {\doibase 10.12942/lrr-2004-4} {\bibfield  {journal}
  {\bibinfo  {journal} {Living Rev. Rel.}\ }\textbf {\bibinfo {volume} {7}},\
  \bibinfo {pages} {4} (\bibinfo {year} {2004})}\BibitemShut {NoStop}%
\end{thebibliography}%
\end{document}